\documentclass[twocolumn]{svjour3}  
\smartqed  

\usepackage[dvipdfmx]{color}
\usepackage{graphicx}
\usepackage[T1]{fontenc}
\usepackage{mathptmx}  
\usepackage[scaled]{helvet}  
\usepackage{courier}
\usepackage[subrefformat=parens]{subcaption}

\makeatletter
\let\cl@chapter\undefined
\makeatletter
\usepackage{amsmath,amssymb}   
\usepackage{mathtools} 
\usepackage{cleveref}  
\Crefname{equation}{Eq.}{Eqs.}%
\Crefname{figure}{Fig.}{Figs.}%
\usepackage{amsbsy}
\usepackage{bm}
\usepackage{algorithmic}
\usepackage{algorithm}
\providecommand{\argmin}{\operatornamewithlimits{argmin}} 

\journalname{Preprint}
\begin{document}

\title{System Parameter Exploration of Ship Maneuvering Model for Automatic Docking / Berthing using CMA-ES 
}

\author{Yoshiki Miyauchi         \and Atsuo Maki \and
        Naoya Umeda              \and Dimas M. Rachman \and
        Youhei Akimoto
}

\institute{Y. Miyauchi \and A. Maki \and N. Umeda \and D. M. Rachman \at
              Osaka University, 2-1 Yamadaoka, Suita, Osaka, Japan \\
              \email{yoshiki\_miyauchi@naoe.eng.osaka-u.ac.jp}  \\
              \email{maki@naoe.eng.osaka-u.ac.jp} 
           \and
           Y. Akimoto \at
           Faculty of Engineering, Information and Systems, University of Tsukuba, 1-1-1 Tennodai, Tsukuba, Ibaraki 305-8573, Japan \\
           RIKEN Center for Advanced Intelligence Project, 1-4-1 Nihonbashi, Chuo-ku, Tokyo 103-0027, Japan
}

\date{Received: date / Accepted: date}

\maketitle

\begin{abstract}
Accurate maneuvering estimation is essential to establish autonomous berthing control. The system-based mathematical model is widely used to estimate the ship's maneuver. Commonly, the system parameters of the mathematical model are obtained by the captive model test (CMT), which is time-consuming to construct an accurate model suitable for complex berthing maneuvers.
System identification (SI) is an alternative to constructing the mathematical model. However, SI on the mathematical model of ship's maneuver has been only conducted on much simpler maneuver: turning and zig-zag. Therefore, this study investigates the SI on a mathematical model capable of berthing maneuver. The main contributions of this study are as follows: (i) construct the system-based mathematical model on berthing by optimizing system parameters with a reduced amount of model tests than the CMT-based scheme; (ii) Find the favorable choice of objective function and type of training data for optimization. Global optimization scheme CMA-ES explored the system parameters of the MMG model from the free-running model's trajectories.  The berthing simulation with the parameters obtained by the proposed method showed better agreement with the free-running model test than parameters obtained by the CMT. Furthermore, the proposed method required fewer data amounts than a CMT-based scheme.
\end{abstract}

\section{Introduction}
\label{sec:intro}
In recent years, the autonomous operation of ships has been actively studied, and autonomous berthing is one of the critical technologies. To develop the autonomous berthing control algorithm, accurate prediction of ship maneuvering by the numerical method is essential because the ship is operated near the berth wall, and estimation error may cause a devastating result. Several methods to predict the ship's maneuver had been used: system-based method, direct estimation by computational fluid dynamics (CFD), and black-box model.

\paragraph{System-based method}
The most commonly used method to estimate the ship's maneuver is the system-based method, which represents the dynamics by several mathematical models consisted of several sets of equations and system parameters (hereafter, maneuvering model and mathematical model will be used interchangeably when referring to the representation of the dynamical system of ship's maneuver). The Mathematical Maneuvering Group (MMG) model~\cite{Ogawa1978} and the Abkowitz model~\cite{Abkowitz1964} are major system-based methods. Both models need to obtain the inherent system parameters of each ship; the captive model test (CMT)~\cite{Yasukawa2015} and empirical formulae~\cite{Sukas2019} are widely used to obtain system parameters. Computational fluid dynamics~\cite{Sakamoto2019a,Guo2017,Guo2018,Guo2020,Villa2019,Muscari2017a,Bhushan2019} is an alternative method to obtain hydrodynamic forces to obtain system parameters. Another method to establish the system-based mathematical model is System Identification (SI), which estimates system parameters from the time history of motion of ship obtained by free-running model tests, full-scale ship trial results, or numerical simulation. Several optimization methods for SI had been reported, both on estimation on MMG model and Abkobitz model: Kalman filter~\cite{Abkowitz1980}; the least-square method~\cite{Araki2012a,Jian-Chuan2015}; support vector regression~\cite{Jian2015,Luo2016,Liu2019,Xu2020a}; generic algorithm~\cite{Sutulo2014,Bonci2015}, and Bayesian approach~\cite{Xue2020}.

\paragraph{Direct CFD}
The second option is estimating all hydrodynamic force and rigid motion simultaneously by viscous CFD. Although this direct CFD method requires intense computational resources, generally, it is independent of hydrodynamic modeling except for turbulence modeling on Reynolds-averaged Navier-Stokes (RANS) equations. Numerous studies are done on direct estimation by CFD, such as turning and zig-zag motion~\cite{Araki2012a,Bonci2015,Carrica2013a,Mofidi2014,Wang2018,Dubbioso2013,Dubbioso2016,Jin2019}, crush astern maneuver~\cite{Wang2020}.
The direct CFD estimation is more capable of accuracy than the system-based method compared with the free-running model test~\cite{Araki2012a}.

\paragraph{Black-box model}
The third option is the black-box model; the model does not have an explicit form of equations and only requires input and output for training. The black-box model has a potentially higher capability to express complex non-linear dynamics than the system-based method because the predetermined structure of mathematical models bounds the capability of the system-based method. Numerous studies had been done in the last decade: recursive neural network (RNN)~\cite{Moreira2003,Oskin2013}; locally weighted learning\cite{Bai2019}; model reference and random forest~\cite{Mei2019}.

When comparing three methods from the perspective of online control, the system-based method with the SI technique is one of the most favorable choices. The direct CFD method is not practical due to its intense computational cost, although it is accurate. In contrast, once the model was established, the system-based method and black-box model are applicable to online control algorithms due to their short computational time. In practice, the system-based method with the CMT for parameter acquisition (hereafter, referred to as ``CMT-based scheme'') have been commonly used in the field of berthing control~\cite{Hasegawa1993,Hasegawa1994,Ahmed2013,Li2020a,Maki2020b}, however, this method has several drawbacks: a large amount of model test is required; difficult to achieve quantitative agreement with free-run model test\cite{Araki2012a}; scale effect caused by difference of Reynolds number between full-scale ship and model. Because of these points, models are needed to be tuned manually by humans before practical use~\cite{Sutulo2014}.
On the other hand, system-based method with SI and black-box model are favorable because they only require few trajectories and are able to avoid the scale effect when the full-scale ship's trajectory is used as input (e.g., ~\cite{Bai2019,Zhu2020}). Although the black-box model can represent complex dynamics better than the system-based method, the dynamics of the black-box model are not understandable, and can not expect the model's behavior when the given control and state are an extrapolation of the training data. Hence, from the safety perspective, the authors consider the system-based method with SI more suitable for developing a maneuvering model for autonomous berthing control. 

However, to the best of our knowledge, the research on SI of maneuvering mathematical model have been done only on turning and zig-zag maneuver. Those are maneuvers on the open sea, which means the ship is advancing all time, the propeller is operating at forwarding direction and constant revolution. Berthing maneuver includes various state and control input combinations: changing control input in a wider range, such as switching the direction of propeller revolution; and ship is forwarding and astern or crabbing. The standard MMG model assumes only maneuvering on the open sea. Hence, additional mathematical models are required to estimate berthing maneuvers. Several sub-model for the MMG model have been proposed to model slow-speed region: e.g., propeller reversal~\cite{Yoshimura1978,Hasegawa1994}; hull force for large drift condition~\cite{KOSE1985,Yoshimura2009a,KOBAYASHI1994}; rudder force~\cite{KOBAYASHI1994,Yasukawa2021}. By introducing these additional sub-models, the MMG model will be able to compute berthing maneuver~\cite{Hasegawa1994,Li2020a,Sawada2020}. However, those ``model-rich'' MMG model requires additional scale model test or numerical simulation to obtain system parameters, which will be costly and time-consuming. From the perspective of time and cost, SI is preferable to facilitate a mathematical model of berthing, which requires only several trajectories of maneuver. Although these demands, we cannot find the research on SI of maneuvering model for berthing, other than research done by the authors' group~\cite{Nishikawa2020}. 

\subsection{Objective and scope}
Consequently, research on the SI of the mathematical model for the berthing is necessary.
This study investigates the feasibility of system parameter exploration on the mathematical model for berthing maneuvers from physically obtained trajectories. The main contributions of this study are as follows: (i) construct the system-based mathematical maneuvering model on berthing by the optimization of system parameters with a reduced amount of model test than CMT-based scheme; (ii) Find the favorable choice of objective function and type of trajectory as input data for optimization of system parameters on berthing maneuver.

The rest of the paper is organized as follows: section two describes the mathematical model of maneuvering; section three describes the optimization scheme used in this study for exploration of system parameters; section four shows the method of data set generation and data set itself; section five shows the results, which are comparison on optimal parameter and Conservative EFD obtained parameter; finally, section six gives the discussion on obtained results and section seven concludes the study.

\section{Mathematical model of ship maneuver}
\label{sec:mmg}
\begin{figure}
    \centering
    \includegraphics[width=0.8\linewidth]{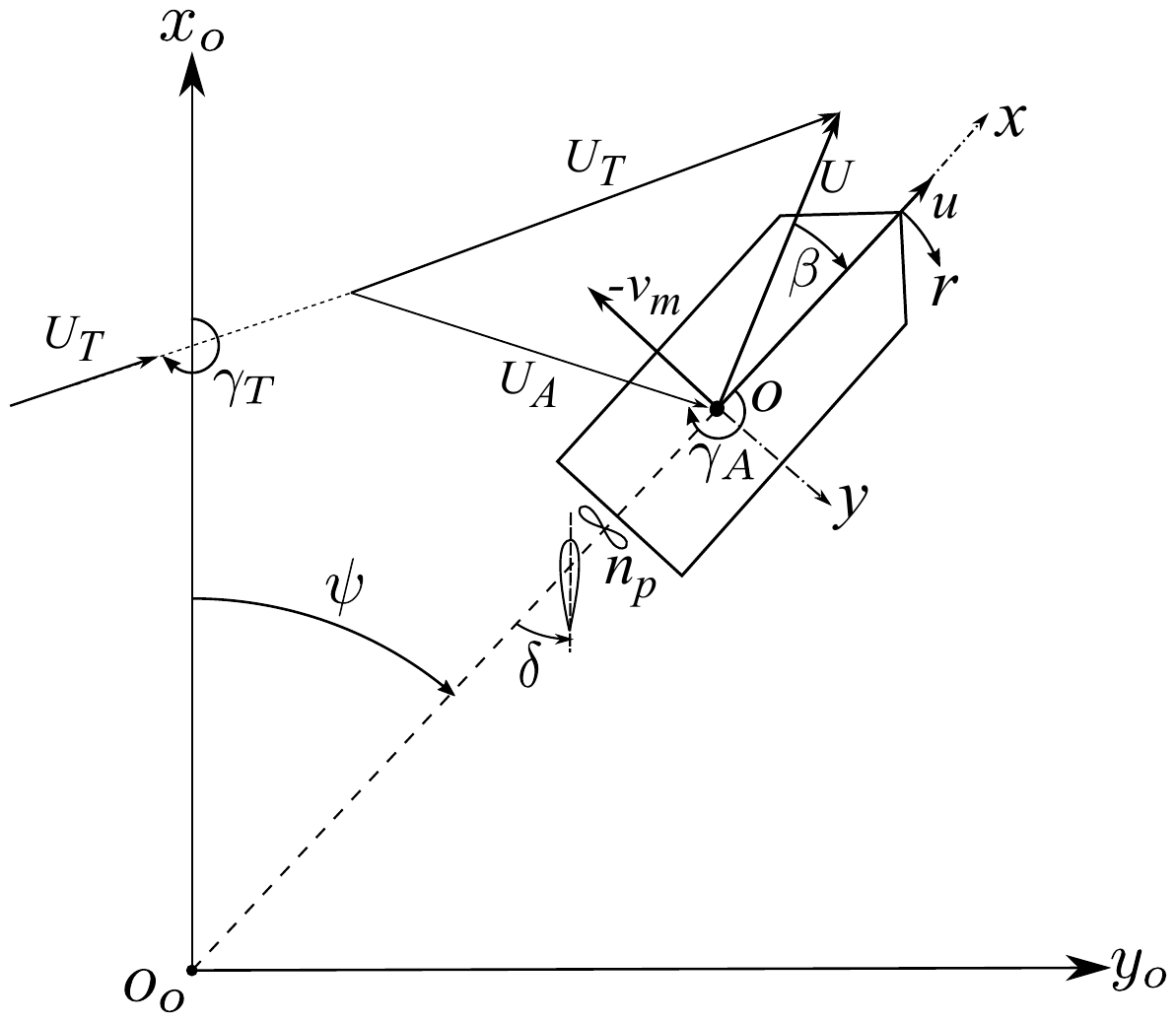}
    \caption{Coordinate System}
    \label{fig:coordinate}
\end{figure}

In this study, the maneuver of the ship was modeled as a 3 degrees-of-freedom problem on the surge, sway, and yaw motion. The coordinate systems are space-fixed system $o_{0}-x_{0}y_{0}$ and ship-fixed system $o-xy$, which has its origin on midship.  \Cref{fig:coordinate} shows the coordinate systems in this study. State vector is $\boldsymbol{x}=(x_{0}, u, y_{0}, v_{m}, \psi, r)^\mathsf{T}  \in \mathbb{R}^{6}$, where $u,v_{m} $ are the velocity of $o-xy$ system. The vector of control input is $\boldsymbol{u}=(\delta,\ n_{p})^\mathsf{T} $, represent the rudder angle, the revolution of propeller respectively. Wind disturbance was considered as environmental parameter $\boldsymbol{\omega}$, which consist by the true wind direction and true wind speed $\boldsymbol{\omega}=(\gamma_{\mathrm{T}}, \ U_{\mathrm{T}})^\mathsf{T} $. However, due to the ship's maneuver, an apparent wind affects the actual force acting on the hull. Hence we computed the apparent wind inside the mathematical model of maneuvering. The zero direction of true wind direction $\gamma_{\mathrm{T}}$ was set to the direction in which the wind blows from the positive direction to the negative direction of $x_{0}$.

The 3-DoF equation of MMG model is express as follows:
\begin{equation}\label{eq:MMGdynamics}
    \begin{split}
    (m + \underline{m_x}) \dot{u} - (m + \underline{m_y}) v_m r -x_{G}mr^{2} &= X\\
    (m + \underline{m_y}) \dot{v}_m + (m + \underline{m_x}) u r +x_{G}m\dot{r} &= Y\\
    (\underline{I_{zz} + J_{zz}}+x_{G}^{2}m) \dot{r} + x_G m (\dot{v}_m + ur) &= N
    \end{split}
\end{equation}
with
\begin{equation}
    \begin{split}
    X &= X_H + X_P + X_R + X_A \\
    Y &= Y_H + Y_P + Y_R  + Y_A\\
    N &= N_H + N_P + N_R + N_A\enspace.
    \label{eq:MMGcompose}
    \end{split}
\end{equation}
Here, the dot (e.g. $\dot{x})$ is the time derivative. The right-hand side of \Cref{eq:MMGdynamics} represents the force or moment acting on the ship, and the MMG model decomposes the hydrodynamic force acting on the ship to sub-model for major components consisting of the ship as \Cref{eq:MMGcompose}. The subscripts H, P, R, and  A denote the hull, the propeller, the rudder, and the external forces by wind, respectively. MMG model has several system parameters $\boldsymbol{\theta}$, and estimate the time derivative $\dot{\boldsymbol{x}}(t)$ by solving $\boldsymbol{f}$ expressed in \Cref{eq:MMGdynamics}: 
    \begin{equation}
        \dot{\boldsymbol{x}}(t) = \boldsymbol{f}\big\{ \boldsymbol{x}(t), \boldsymbol{u}(t), \boldsymbol{\omega}(t); \ \boldsymbol{\theta} \big\} \enspace.
        \label{eq:dynamics}
    \end{equation}
In this study, we show the method to find optimal $\boldsymbol{\theta}$. Hereafter, parameters with underline (e.g., $\underline{m_x}$) means those parameters were explored in the optimization process. In the next section, the details of each sub-models are explained. 

\subsection{Force on Hull}\label{sec:hull}
The force acting on the hull was computed by the unified model for navigation on the open sea and harbor maneuvers~\cite{Yoshimura2009a}:
\begin{align}\label{eq:yoshimura}
    \begin{aligned}
        X_{H}=&\left(\frac{\rho}{2}\right) L_{\mathrm{pp}} d
        \left[ \begin{array}{ll}\begin{split}
            &\left\{X_{0(F)}^{\prime}+\left(\underline{X_{O(A)}^{\prime}}-X_{0(F)}^{\prime}\right)(\beta / \pi)\right\} u U \\
            &+\underline{X_{v r}^{\prime}} L_{\mathrm{pp}} \cdot v_{m} r
            \end{split}\end{array}\right] \\
        Y_{H}=&\left(\frac{\rho}{2}\right) L_{\mathrm{pp}} d\\
        &\quad \left[ \begin{array}{ll}\begin{split}
            &\underline{Y_{v}^{\prime}} v_{m}|u|+\underline{Y_{r}^{\prime}} L_{\mathrm{pp}} \cdot r u \\
            &-\left(\frac{\underline{C_{D}}}{L_{\mathrm{pp}}}\right) \int_{-L_{\mathrm{pp}} / 2}^{L_{\mathrm{pp}} / 2}\left|v_{m}+\underline{C_{rY}} r x\right|\left(v_{m}+\underline{C_{rY}} r x\right) d x
        \end{split}\end{array}\right] \\
        N_{H}=&\left(\frac{\rho}{2}\right) L_{\mathrm{pp}}^{2} d \\
        &\quad \left[ \begin{array}{ll}\begin{split}
            &\underline{N_{v}^{\prime}} v_{m} u+\underline{N_{r}^{\prime}} L_{\mathrm{pp}} \cdot r|u| \\
            &-\left(\frac{\underline{C_{D}}}{L_{\mathrm{pp}}^{2}}\right) \int_{-L_{\mathrm{pp}} / 2}^{L_{\mathrm{pp}} / 2} \left|v_{m}+\underline{C_{rN}} r x \right| \left(v_{m}+\underline{C_{r N}} r x\right) x d x
        \end{split}\end{array}\right]\enspace,
    \end{aligned}
\end{align}
where $\rho,$ density of water; $L_{\mathrm{pp}}$, Length between perpendiculars of ship; $d$, draft of ship; $X_{O(F)}^{\prime} \text{and} \ X_{O(A)}^{\prime}$, resistance coefficients of ahead and astern; $C_{D}$, cross flow drag coefficient; $C_{r Y} \ \text{and } C_{r N}$, correction factor for lateral force and yaw moment; $X_{0(F)}^{\prime}, \ Y_{v}^{\prime}, \ Y_{r}^{\prime}, \ N_{v}^{\prime}, \ \text{and}\ N_{r}^{\prime}$ are non-demensional hydrodynamic derivatives, respectively. Hereafter, superscript prime (e.g.\ $ X_{0(F)}^{\prime})$ means the non-dimensionalized value. Note that added mass term in original expression of \Cref{eq:yoshimura} was move to left hand side of \Cref{eq:MMGdynamics}.

\subsection{Force by propeller}
Since standard MMG model~\cite{Yasukawa2015} assumed only the forwarding maneuver $(u>0,\ n_{\mathrm{p}}>0)$, computation on propeller force applied  additional sub-models based on operation condition of propeller, which are divided by quadrant: first $(u\geq0,\ n_{\mathrm{p}}>0)$; second $(u<0,\ n_{\mathrm{p}}>0)$; third $(u\geq0,\ n_{\mathrm{p}}<0)$; and fourth $(u<0,\ n_{\mathrm{p}}<0)$. On the first and second quadrant, propeller thrust was computed by standard MMG model:
\begin{equation}
    X_{P}=\rho n_{p}^{2} D_{p}^{4}\left(1-\underline{t_{p}}\right) K_{T}\enspace,
\end{equation}
where thrust coefficient $K_{T}$ was express by a pronominal expression of advance coefficient $J_{p}=\left(1-w_{p}\right) u /\left(n_{p} D_{p}\right)$.
The effective propeller wake fraction $w_{p}$ was computed as follows~\cite{ITTC2002}:
\begin{equation}
    1-w_{p}=1-\underline{w_{p 0}}+\underline{\tau}\left|v_{m}^{\prime}+\underline{x_{p}^{\prime}} r^{\prime}\right|+\underline{C_{P}^{\prime}}\left(v_{m}^{\prime}+\underline{x_{p}^{\prime}} r^{\prime}\right)^{2} \enspace,
\end{equation}
where: $w_{p0}$ is the wake fraction on $v_{m}=r=0$; $\tau, \ C^{\prime}_{P} \text{and} \ x^{\prime}_{p}$ are empirical coefficients. The trust deduction factor $t_p$ and wake fraction $w_{p}$ varies by propeller operation condition, however in this study simply modeled as:
\begin{align}
    t_{p}=0 \quad &\text{for } n_{\mathrm{p}}<0\\
    w_{p}=0 \quad &\text{for } u<0 \enspace,
\end{align}
as model in~\cite{KOBAYASHI1994,Yasukawa2003}. lateral force and yaw moment induced by propeller on first and second quadrant $(n_{\mathrm{p}}\geq 0)$ are usually neglected in MMG model, however in this study , those were computed by the polynomial based on the captive test of training vessel~\cite{Ueno2001}, as follows:
\begin{align}
    Y_{P} = \begin{cases}
    0 &\text{for } u\geq0\\
    \frac{1}{2} \rho L_{\mathrm{pp}}^{2} d \left(n_{p} P\right)^{2}\left(\underline{A_{6}} J_{s}^{2}+\underline{A_{7}} J_{S}+\underline{A_{8}}\right) &\text{for } u<0
    \end{cases}\\
    N_{P} = \begin{cases}
    0 &\text{for } u\geq0\\
    \frac{1}{2} \rho L_{\mathrm{pp}}^{2} d \left(n_{p} P\right)^{2}\left(\underline{B_{6}} J_{s}^{2}+\underline{B_{7}} J_{S}+\underline{B_{8}}\right) &\text{for } u<0
    \end{cases}
\end{align}
where: $P$, pitch of propeller; $J_{s}=u/(n_{\mathrm{P}}D_{\mathrm{P}})$; $A_{6} \ \text{through} \ A_{8}$ and $B_{6} \ \text{through} \ B_{8}$ are polynomial coefficients.

On the propeller reversal condition, same as second quadrant, polynomial expression~\cite{Hasegawa1994} based on CMT was used:
\begin{align}
    X_{P} =& \rho n_{p}^{2} D_{p}^{4}\begin{cases}
        \underline{C_{6}}+\underline{C_{7}} J_{s} & \text{for~} \left(J_{s} \geq \underline{C_{10}}\right) \\
        \underline{C_{3}} & \text{for~} \left(J_{s}<\underline{C_{10}}\right)
    \end{cases} \\
    Y_{p}=&\frac{1}{2} \rho L d\left(n_{p} D_{p}\right)^{2}
    \begin{cases}
        \underline{A_{1}}+\underline{A_{2}} J_{s} & \left(-0.35 \leq J_{s} \leq -0.06\right) \\
        \underline{A_{3}}+\underline{A_{4}} J_{s} & \left(J_{s}<-0.35\right) \\
        \underline{A_{5}} & \left(-0.06<J_{s}\right)
    \end{cases}\\
    N_{p}=&\frac{1}{2} \rho L^{2} d\left(n_{p} D_{p}\right)^{2}\begin{cases}
        \underline{B_{1}}+\underline{B_{2}} J_{s} & \left(-0.35 \leq J_{s} \leq -0.06\right) \\
        \underline{B_{3}}+\underline{B_{4}} J_{s} & \left(J_{s}<-0.35\right) \\
        \underline{B_{5}} & \left(-0.06<J_{s}\right) \enspace,    
    \end{cases}
\end{align}
where, $A_{1} \ \text{through} \ A_{5}$, $B_{1} \ \text{through} \ B_{5}$, $C_{3}, \ C_{6}, \ C_{7},$ and $C_{10}$ are polynomial coefficients.

\subsection{Force by Rudder}
Induced force and moment by rudder were expressed as follows on standard MMG model~\cite{Yasukawa2015}:
\begin{align}
    X_{R}&=-\left(1-\underline{t_{R}}\right) F_{N} \sin \delta \\
    Y_{R}&=-\left(1-\underline{a_{H}}\right) F_{N} \cos \delta \\
    N_{R}&=-\left(x_{R}+\underline{a_{H}}\ \underline{x_{H}}\right) F_{N} \cos \delta \enspace,    
\end{align}
where $F_{N}$ is the rudder normal force:
\begin{equation}
    F_{N}=(1 / 2) \rho A_{\mathrm{R}} U_{\mathrm{R}}^{2} f_{\alpha} \sin \alpha_{\mathrm{R}}\enspace.
\end{equation}
Here, $t_{R}$, thrust deduction factor by steering; $x_{R}$, longitudinal position of the rudder from midship; $a_{H}$, correction factor lateral force; $x_{H}$ position of additional lateral force on hull due to steering; $A_{R}$, area of the rudder. The gradient of rudder normal force $f_{\alpha}$ is expressed as constant in most MMG model-related studies, however, regarding the berthing maneuver, a rudder operates even in the deep-stall region; hence $f_{\alpha}$ is not constant anymore. Even so,  $f_{\alpha}$ was assumed as constant and computed by the most commonly used empirical expression, Fujii's formula~\cite{Fuji1961}, which is the function of rudder aspect ratio $\lambda$:
\begin{equation}
    f_{\alpha} = 6.13\lambda/(2.25+\lambda)\enspace.
\end{equation}
The resultant rudder inflow speed $U_{R}$ and effective inflow angle $\alpha_{R}$ were expressed by longitudinal and lateral inflow speed $u_{R}$ and $v_{R}$:
\begin{align}
    U_{\mathrm{R}}&=\sqrt{u_{\mathrm{R}}^{2}+v_{\mathrm{R}}^{2}} \\
    \alpha_{\mathrm{R}}&=\delta-\mathtt{atan2}\left(\frac{v_{\mathrm{R}}}{u_{\mathrm{R}}}\right) \enspace.\label{eq:atan2}
\end{align}
We extended the standard MMG model to \Cref{eq:atan2} to apply to the berthing maneuver by introducing the function $\mathtt{atan2}(y/x)$ reruns the $\tan^{-1}(y/x)$ in range of $(-\pi, \pi]$. The lateral inflow speed $v_{R}$ is expressed as follows by using flow straightening coefficient $\gamma$ and experimental constant $l_{R}$:
\begin{equation}
v_{R}=\begin{cases}
    -\underline{\gamma_{P}}\left(v_{m}+\underline{l_{R}} r\right) & \text{for } v_{m}+x_{R} r \geq 0\ \\
    -\underline{\gamma_{N}}\left(v_{m}+\underline{l_{R}} r\right) & \text{for } v_{m}+x_{R} r < 0 \enspace .
\end{cases}
\end{equation}
The longitudinal inflow $u_{R}$ will be heavily affected by direction of ship motion and propeller induced flow. For $n_{\mathrm{p}}\geq0$, $u_{R}$ was expressed by modified form~\cite{Yoshimura1978} for low speed region as:
\begin{equation}
\begin{split}
    &u_{R}=\\
    &\underline{\varepsilon} \sqrt{\eta\left\{u_{P}+\frac{\underline{k_{x}}}{\underline{\varepsilon}}\left(\sqrt{u_{P}^{2}+\frac{8 K_{T}\left(n_{P} D_{P}\right)^{2}}{\pi}}-u_{P}\right)\right\}^{2}+(1-\eta) u_{P}^{2}} \enspace.
\end{split}
\end{equation}
Here, $u_{P}=(1-w_{P})u$; $\eta=D_{\mathrm{P}}/H_{R}$; $H_{R}$, height of rudder; $\varepsilon$, ratio of wake fraction; $k_{x}$, empirical coefficient. On the third quadrant, Kitagawa's model~\cite{Kitagawa2015} was applied:
\begin{equation}
u_{R}=\operatorname{sgn}\left(u_{R s q}\right) \cdot \sqrt{\left|u_{R s q}\right|} \enspace,
\end{equation}
where:
\begin{align}
 \begin{split}
   u_{R s q}=&\eta \cdot \operatorname{sgn}\left(u_{R P R 1}\right) \cdot u_{R P R 1}^{2}\\
   &\quad +(1-\eta) \operatorname{sgn}\left(u_{R P R 2}\right) \cdot u_{R P R 2}^{2}+\underline{C_{PR}} \cdot u
\end{split}\\
    u_{R P R 1}&=u \varepsilon\left(1-w_{p}\right)+n_{p} D_{p} \underline{k_{x PR}} \sqrt{8\left|K_{T}\right| / \pi} \\
    u_{R P R 2}&=u \varepsilon\left(1-w_{p}\right) \enspace.
\end{align}
Here, $k_{xPR}$ and $C_{PR}$ are the velocity increase factor and the correction factor for propeller reversal condition, respectively.
On the fourth quadrant, we assumed that the inflow is equal to the ship's motion: $u_{R}=u$~\cite{KOBAYASHI1994}.
\subsection{Force by Wind}

Regarding the external force induced by wind disturbance, Fujiwara's regression formulae~\cite{Fujiwara1998} was used to estimate the wind pressure coefficients:
\begin{align}
    \begin{aligned}
        X_{A} &= (1/2)\rho_{A}U_{A}^{2}A_{T}\cdot C_{X} \\
        Y_{A} &= (1/2)\rho_{A}U_{A}^{2}A_{L}\cdot C_{Y} \\
        N_{A} &= (1/2)\rho_{A}U_{A}^{2}A_{L}L_{OA}\cdot C_{N} \enspace,
    \end{aligned}
\end{align}
where
\begin{align}
    \begin{aligned}
        C_{X} =& \underline{X_{0}}+\underline{X_{1}} \cos (2\pi - \gamma_{A})+X_{3} \cos 3 (2\pi - \gamma_{A}) \\
              &+\underline{X_{5}} \cos 5 (2\pi - \gamma_{A}) \\
        C_{Y} =& \underline{Y_{1}} \sin (2\pi - \gamma_{A})+\underline{Y_{3}} \sin 3 (2\pi - \gamma_{A}) \\
              &+\underline{Y_{5}} \sin 5 (2\pi - \gamma_{A}) \\
        C_{N} =& \underline{N_{1}} \sin (2\pi - \gamma_{A})+\underline{N_{2}} \sin 2 (2\pi - \gamma_{A}) \\
              &+\underline{N_{3}} \sin 3 (2\pi - \gamma_{A}) \enspace.
    \end{aligned}
\end{align}
Here, $\rho_{A}$ is the density of air, $A_{T},\ A_{L}, L_{OA}$ are the transverse projected area, the lateral projected area, and the overall length of the ship, respectively. $X_{i}, \ Y_{i}, N_{i}$ are coefficients to express wind pressure coefficients derived by the regression formulae~\cite{Fujiwara1998} which use geometric parameters of the ship as explanatory variables and based on wind tunnel test data of numerous scaled ship models.

\subsection{EFD model}\label{sec:efdmodel}
The common method to obtain system parameters of the MMG model is a test using the scaled model, such as a CMT at the towing tank facility. To evaluate the optimal parameters obtained by the proposed method, the MMG model with parameters obtained by model test and empirical formulae (hereafter, referred to as ``Experimental fluid dynamics (EFD) model'') was used as a reference. \Cref{tab:params} shows the list of parameters which was optimized in this study and the source of parameters of the EFD model. Most of the parameters were obtained by the CMT of the subject ship; however, several parameters were substituted by empirical formulae or other ship's values due to the absence of data on the subject ship.
\begin{table}[tb]
    \centering
        \caption{List of parameters optimized on this study and source of EFD model.}
    \begin{tabular}{ll}
        \hline parameters &  Source\\
         \hline\hline
         $m_{x}$, $m_{y}$, $I_{zz}+J_{zz}$& Empirical formulae \\
         $X_{O(A)}^{\prime}$,$X_{vr}^{\prime}$,& Captive test~\cite{Hachii2004}\\
         $Y_{v}^{\prime}$, $Y_{r}^{\prime}$, $N_{v}^{\prime}$, $N_{r}^{\prime}$\\
         $C_{D}$, $C_{rY},C_{rN}$ & Empirical formulae~\cite{Yoshimura2009a}\\
         $t_{P}$, $w_{P0}$, $\tau$, $x_{P}^{\prime}$, $C_{P}^{\prime}$ & Captive test~\cite{ITTC2002}\\
         $A_{1}$,$A_{2}$,$A_{3}$,$A_{4}$,$A_{5}$ & Captive test~\cite{Hasegawa1994}\\
         $B_{1}$,$B_{2}$,$B_{3}$,$B_{4}$,$B_{5}$ & \\
         $C_{3}$, $C_{6}$, $C_{7}$, $C_{10}$ & \\
         $A_{6}$, $A_{7}$, $A_{8}$, & Captive test of train vessel~\cite{Ueno2001}\\
         $B_{6}$, $B_{7}$, $B_{8}$  &\\
         $t_{R}$, $a_{H}$, $x_{H},$  &Captive test~\cite{Hachii2004}\\
         $\gamma_{P}$, $\gamma_{N}$, $l_{R}$, $k_{x}$, $\varepsilon$ & \\
         $k_{xPR}$, $C_{PR}$ & Captive test of KVLCC1 and Bulk Carrier~\cite{Kitagawa2015}\\
         $X_{0}$, $X_{1}$, $X_{3}$, $X_{5}$ & Regression formulae~\cite{Fujiwara1998}\\
         $Y_{1}$, $Y_{3}$, $Y_{5}$, &\\
         $N_{1}$, $N_{2}$, $N_{5}$ &\\
         \hline
    \end{tabular}
    \label{tab:params}
\end{table}

\section{Optimization scheme}
\subsection{Objective Function}\label{sec:J}
The optimization of MMG model was defined as minimization problem on the difference of maneuver between input data set $\mathcal{D}$ and simulation using the obtained numerical model. In this study, $\mathcal{D}$ consists with several trajectories measured on free-run model test (i.e. turning, zig-zag, and random maneuver), which includes time history of model test: $\boldsymbol{x}_{\text{input}}(t), \ \boldsymbol{u}_{\text{input}}(t)$ and $\boldsymbol{\omega}_{\text{input}}(t)$. Trajectories contained in $\mathcal{D}$ were divided to contiguous subsequence to mitigate the effect of error accumulation of maneuvering simulation on the optimization process. The simulation using the MMG model estimates the maneuver as an initial value problem for each contiguous subsequence  of $\mathcal{D}$: 
\begin{align}
    \boldsymbol{x}_{\text{input}}(0)&=\boldsymbol{x}_{\text{sim}}(0) \\ \boldsymbol{u}_{\text{input}}(t)&=\boldsymbol{u}_{\text{sim}}(t) \\ \boldsymbol{\omega}_{\text{input}}(t)&=\boldsymbol{\omega}_{\text{sim}}(t) \enspace.
\end{align}
Contiguous subsequence is the portion of $\mathcal{D}$ with constant duration. Duration of contiguous subsequence  is $t_{\mathrm{f}}=100$ s, except for final contiguous subsequence  of each trajectory. The period of 100 s is a typical period of berthing maneuver, which is roughly equivalent to 17 minutes on the full scale. On the numerical simulation, the 4th order Runge-Kutta method was used for the time development of \Cref{eq:dynamics}.

The optimization of MMG model was formulated as exploration of the optimal parameter vector $\boldsymbol{\theta}_{\text{opt}}$ from domain $\Theta$ which minimize the objective function $J$ on whole data set $\mathcal{D}$:
\begin{align}\label{eq:j}
    \begin{aligned}
        &\boldsymbol{\theta}_{\text{opt}} =\argmin_{\boldsymbol{\theta} \in \Theta} \ J(\theta ;\ \mathcal{D})\\
        &\text { where } \ J \equiv \sum_{i=1}^{N} \int_{0}^{t_{\mathrm{f}}} \| \boldsymbol{\hat{z}}^{i}_{\textrm{input} }(t)-\boldsymbol{\hat{z}}^{i}_{\textrm{sim}}(t) \|^{2} d t \enspace .
    \end{aligned}
\end{align}
Here, $\boldsymbol{\Hat{z}}^{i}(t)=(\Hat{z}^{i}_{1}(t),\ \Hat{z}^{i}_{2}(t), \cdots \Hat{z}^{i}_{j}(t))$ is standardized state vector:
\begin{align}
    \Hat{z}^{i}_{\text{input},j}(t)&=\left(z_{\text{input},j}^{i}(t)-\mu_{\text{input},j}^{i} \right) / \sigma_{\text{input},j}^{i} \\
    \Hat{z}^{i}_{\text{sim},j}(t)&=\left(z_{\text{sim},j}^{i}(t)-\mu_{\text{sim},j}^{i} \right) / \sigma_{\text{sim},j}^{i}\enspace ,    
\end{align}
where superscript $i$ means the $i$-th contiguous subsequence  in $\mathcal{D}$; $t_{\mathrm{f}}$ is the time duration of contiguous subsequence ; $N$ is total number of contiguous subsequence  in $\mathcal{D}$; subscript \textit{input} and \textit{sim} mean the input data and numerical simulation of MMG model; $\boldsymbol{z}(t)$ is the state used in optimization process; $\boldsymbol{\mu}^{i}$ and $\boldsymbol{\sigma}^{i}$ are the mean and standard deviation of $\boldsymbol{z}^{i}(t)$. Detail of $\Theta$ are described on \Cref{sec:domain}.

On the choice of component of the state as input, several choices of $\boldsymbol{z}(t)$ can be taken. The authors previously used velocity component $\boldsymbol{z}(t)= \left(u, \ v_{m},\ r\right)^{\mathsf{T}}$~\cite{Nishikawa2020}. Other options are: to contain both trajectory and velocity as~\cite{Sutulo2014}; or use the representative value of trajectory such as tactical diameter on turning and overshoot angle on the zig-zag test as~\cite{Bonci2015}. Regarding the estimation of berthing maneuver, estimation of location and velocity are both important because berthing maneuver needs to stop at the designated berthing point precisely without collision to berth wall, at zero speed. Hence, the following three options of  $\boldsymbol{z}(t)$ on optimization were compared: 
\begin{align}\label{eq:z1}
    \boldsymbol{z}_{1}^{i}(t) \equiv & \Big(u^{i}(t), \ v_{m}^{i}(t), \ r^{i}(t)\Big)^{\mathsf{T}} \in \mathbb{R}^{3} \\
    \begin{split}\label{eq:z2}
        \boldsymbol{z}_{2}^{i}(t) \equiv & \Big(x_{0}^{i}(t), \ u^{i}(t), \\
        &  y_{0}^{i}(t), \ v_{m}^{i}(t), \ \sin\psi^{i}(t), \ \cos\psi^{i}(t),\ r^{i}(t)\Big)^{\mathsf{T}} \in \mathbb{R}^{7}
    \end{split}\\\label{eq:z3}
    \boldsymbol{z}_{3}^{i}(t) \equiv &\Big(x_{0}^{i}(t), \ y_{0}^{i}(t), \ \sin\psi^{i}(t), \ \cos\psi^{i}(t)\Big)^{\mathsf{T}} \in \mathbb{R}^{4} \enspace.
\end{align}
Hereafter, Objective function which use $\boldsymbol{z}_{1}^{i}(t), \ \boldsymbol{z}_{2}^{i}(t), \ \boldsymbol{z}_{3}^{i}(t)$ called as $J_1, J_2, J_3$, respectively.

\subsection{CMA-ES}
\begin{figure}
    \centering
    \includegraphics[width=\linewidth]{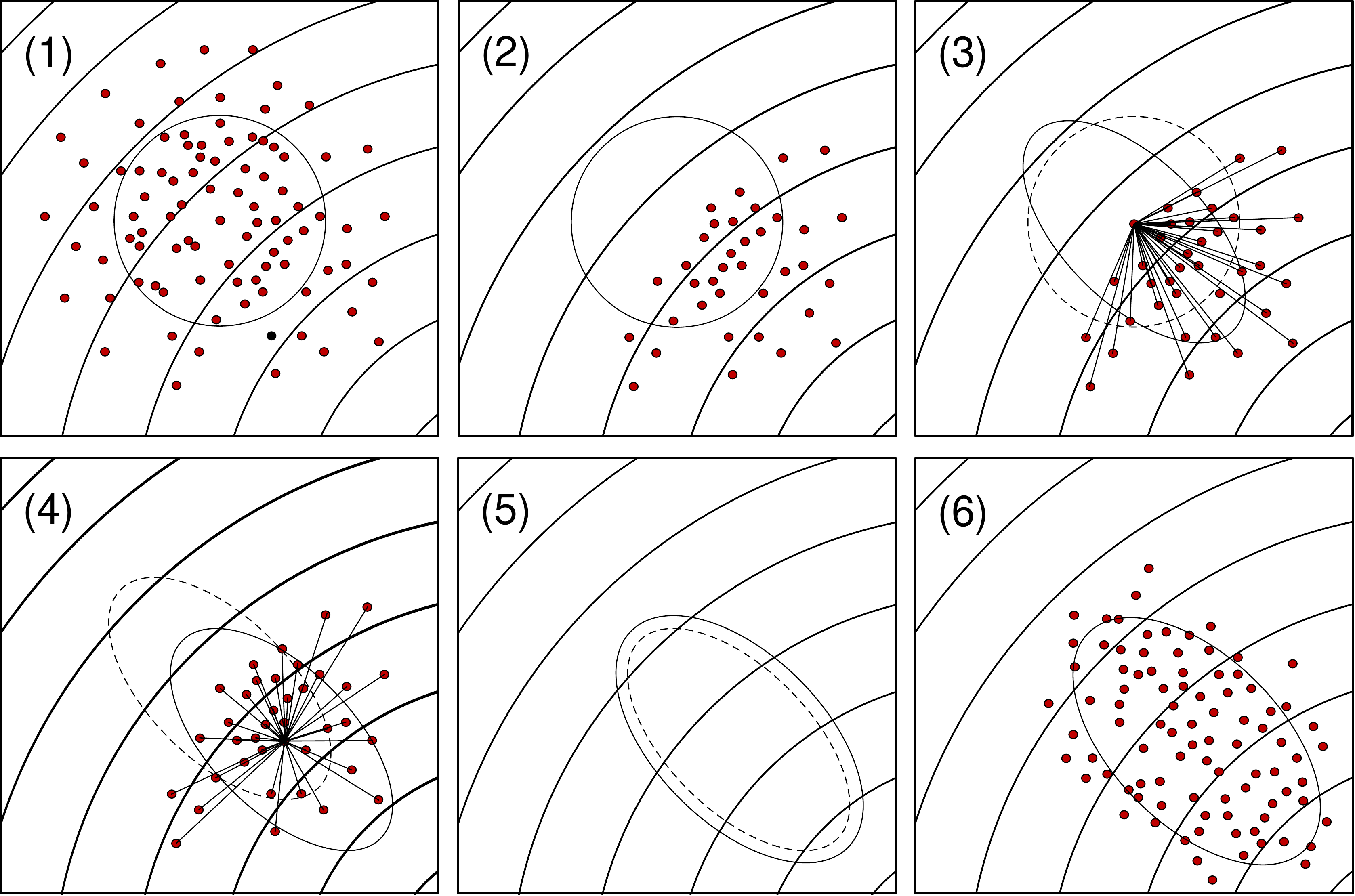}
    \caption{Schematic presentation of the CMA-ES procedure including (1) generating multiple candidate solutions, (2) evaluating and ranking the solutions based on the objective function, (3) updating the covariance matrix, (4) shifting the center of the distribution to a weighted mean vector, (5) updating the step size and (6) generating multiple candidates in the next step. This figure duplicates Fig.~2 in the literature~\cite{Maki2020b}}
    \label{fig:cma}
\end{figure}

In the previous study~\cite{Nishikawa2020,Maki2019a,Maki2019b}, covariance matrix adaption evolution strategy (CMA-ES)~\cite{Hansen2006} with modified box constraints~\cite{Sakamoto2017} and restart strategy~\cite{Auger2005} was applied as the optimization method. Fig.\ref{fig:cma} shows the schematic view of the optimization procedure using CMA-ES. In this study, the initial population size of CMA-ES was set to 20, and the max size was 720, while the population size was doubled when the restart occurred.

\subsection{Range of Parameter Exploration}\label{sec:domain}
Here we show the detail of parameter exploration on this study. Total 57 parameters were explored as shown on \Cref{sec:mmg}. On the optimization by CMA-ES with box-constraint, maximum and minimum value of parameters are required. The $j$-th parameter $\theta_{j}$ was explored within the domain $\Theta$ defined by parameter's value of EFD model $\theta_{\text{EFD},j}$:
\begin{equation}\label{eq:rangeall}
    \theta_{j} \in \Theta_{j} =\big[-10 |\theta_{\text{EFD},j}|, \ 10|\theta_{\text{EFD},j}|\big] \enspace,
\end{equation}
with exception of :
\begin{align}
    \theta_{j} \in
    \begin{cases}
    \big[0.7 \theta_{\text{EFD},j}, \ 1.3\theta_{\text{EFD},j}\big] &\text{for: } m_{x}, \ m_{y}, \ I_{zz}+J_{zz}\\
    \big[10 \theta_{\text{EFD},j}, \ 0.1\theta_{\text{EFD},j}\big] &\text{for: } Y_{v}', \ N_{r}'\\
    \big[0.1 \theta_{\text{EFD},j}, \ 10\theta_{\text{EFD},j}\big] &\text{for: } t_{\mathrm{p}}, \ w_{\mathrm{p}0} \enspace.
    \end{cases}
\end{align}
Those exceptions were made because: (1) added masses affect all the component of force as shown on \Cref{eq:MMGdynamics}, hence broad range of exploration may lead to numerical instability; (2) sign of resistance, $t_{\mathrm{p}}, \ w_{\mathrm{p}0}$ are obvious. Note that $Y_{v}' \ \text{and} \ N_{r}'$ are negative. In this study we used the given parameter $\boldsymbol{\theta_{\text{EFD}}}$, however even if those EFD obtained parameter are not available, for instance, when focusing on newly designed ship, \Cref{eq:rangeall} is broad enough to cover the parameter space with given parameter used in this study.

On the other hand, broad range of exploration may cause the divergence of numerical simulation of maneuvering due to the unrealistically large or small value of the parameters. To maintain numerical stability on optimization process, treatment shown on \Cref{alg:limit} was implemented when $\boldsymbol{\dot{x}}(t)$ exceeded the limit $\boldsymbol{\dot{x}}_{\text{lim}}$. Here, $\boldsymbol{\dot{x}}_{\text{lim}}=(a,a,a,a,a/(0.5L_{\mathrm{PP}}),a/(0.5L_{\mathrm{PP}}))$ where $a=1\times10^{10}$.
\begin{figure}[tb]
\begin{algorithm}[H]
    \caption{Limitation of $\boldsymbol{\dot{x}}(t)$ }
    \label{alg:limit}
    \begin{algorithmic}[1]
    \FOR{$t =0:t_{f}$}
        \STATE get $\boldsymbol{\dot{x}}(t)$ by solving MMG model
        \FOR{k=1:6}
            \IF{$|\dot{x}_{k}(t)|>\dot{x}_{\text{lim},k}$}
                \IF{k=1,3,5}
                \STATE $\dot{x}_{k}(t) =  \text{sgn}\{\dot{x}_{k}(t)\} \{ 2-t/t_{\mathrm f}\} \dot{x}_{\text{lim},k}$  
                \ELSE 
                \STATE $\dot{x}_{k}(t) = \text{sgn}\{\dot{x}_{k}(t)\} \dot{x}_{\text{lim},k}$ 
                \ENDIF
            \ENDIF
        \ENDFOR
    \STATE $\boldsymbol{x}(t+1)$ = $\boldsymbol{x}(t) + \Delta t \boldsymbol{\dot{x}}(t)$
    \ENDFOR
    \end{algorithmic}
\end{algorithm}
\end{figure}

\section{Free-run model test for data set generation}
\subsection{Scale model Ship and instruments}
\begin{figure}[htb]
    \centering
    \includegraphics[width=0.8\columnwidth]{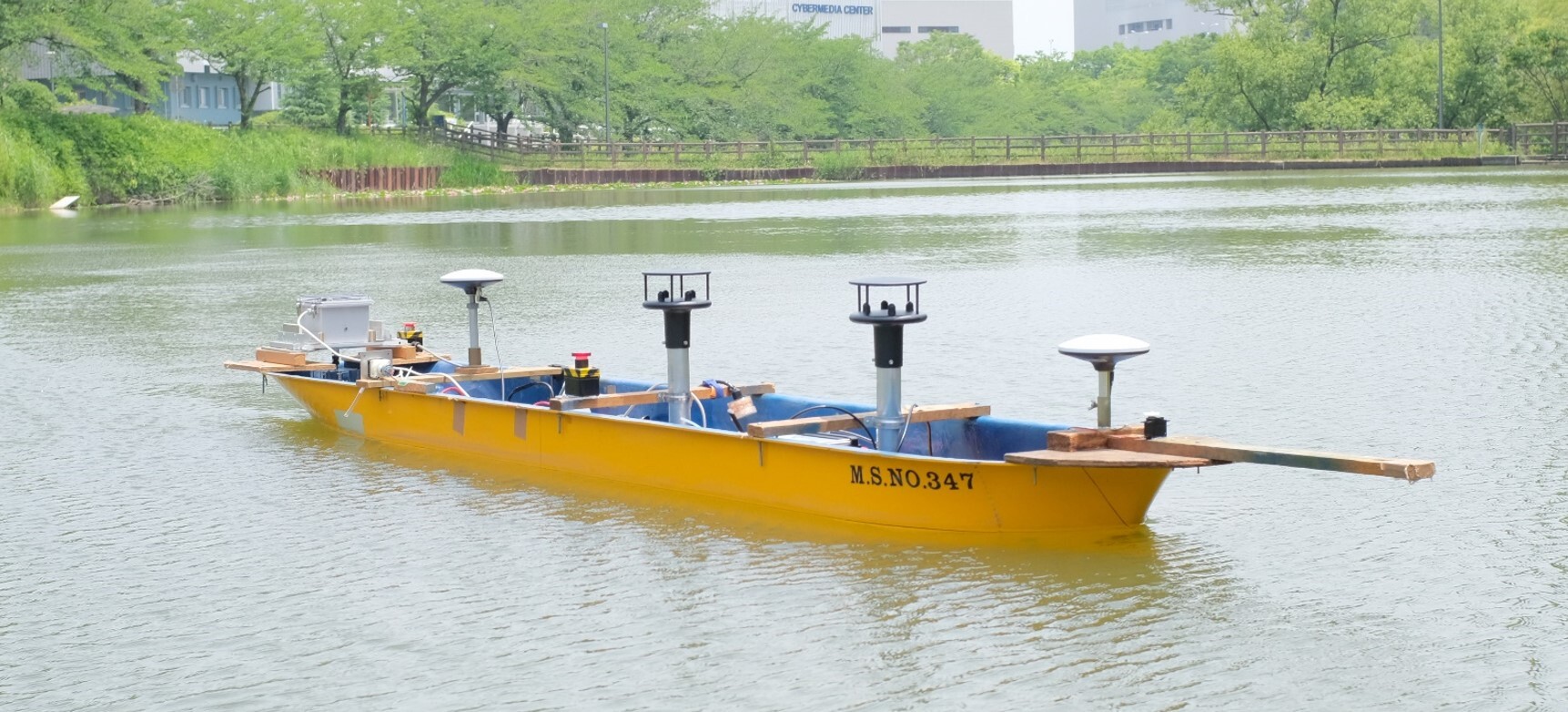}
    \caption{Scale Model Ship of \textit{Esso Osaka}}
    \label{fig:esso_pthoto}
\end{figure}

Training and test data sets were generated by free-run model tests in the experimental pond facility (the \textit{Inukai} pond) at Osaka University using the model ship of VLCC M.V. \textit{Esso Osaka}. \Cref{tab:pp} shows the principal particulars of the model ship. The loading condition is equivalent to the trail condition~\cite{Crane1979}. The model ship is equipped with measurement instruments: a fiber optical gyro (FOG); three GNSS receivers (MJ-3021-GM4-QZS-EVK by Magellan Systems Japan); and two ultrasonic anemometers (Gill PGWS-100-3). From the measured data from these instruments, the time series of state $\boldsymbol{x}(t)$ and true wind speed and direction $\big(U_{T}(t),~ \gamma_{T}(t)\big)^{\mathsf{T}}$ were Estimated. The appearance of the model ship is shown in \Cref{fig:esso_pthoto}.

\begin{table}[tb]
    \centering
        \caption{Principal particulars of subject ship \textit{Esso Osaka}.}
        \begin{tabular}{ll}
        \hline Item & Value \\
        \hline Length between perpendicular: $L_{\mathrm{pp}}(\mathrm{m})$ & $3.0$ \\
        Ship breadth: $B(\mathrm{m})$ & $0.489$ \\
        Ship draft: $d(\mathrm{m})$ & $0.201$ \\
        Diameter of propeller: $D_{\mathrm{p}}(\mathrm{m})$ & $0.084$ \\
        Area of Rudder: $A_{\mathrm{R}}\left(\mathrm{m}^{2}\right)$ & $0.0106$ \\
        Diameter of bow thruster: $D_{\mathrm{BT}}:(\mathrm{m})$ & $0.050$ \\
        Diameter of stern thruster: $D_{\mathrm{ST}}:(\mathrm{m})$ & $0.050$ \\
        Mass: $m$ (kg) & 244.6 \\
        Longitudinal center of gravity: $x_{\mathrm{G}}(\mathrm{m})$ & $0.094$ \\
        Transverse projected area: $A_{\mathrm{T}}\left(\mathrm{m}^{2}\right)$ & $0.135$ \\
        Lateral projected area: $A_{\mathrm{L}}\left(\mathrm{m}^{2}\right)$ & $0.520$ \\
        Block coefficient: $C_{\mathrm{b}}$ & $0.830$ \\
        \hline
        \end{tabular}
    \label{tab:pp}
\end{table}

The details of the measurement and data processing methods are described below. All measurements were performed at 10 Hz. The model ship's trajectory $x_{0}(t), \ y_{0}(t)$ was converted from the GNSS receivers' trajectories to the midship position. The GNSS receivers are compatible with the centimeter-class positioning augmentation service (CLAS). Using CLAS, centimeter-class measurements are possible for moving objects, but the experimental pond facility is an adverse condition for GNSS because surrounded by buildings and other obstructions. Therefore, we monitored the distance between the two GNSS receivers during measurement. The accuracy of GNSS positioning was ensured by using only the measurement results that are less than 5 cm difference from the actual value of the distance between the receivers. The velocity was estimated by numerically differentiating the converted midship trajectory and smoothing it out using a linear Kalman filter. The second-order central difference method was used for numerical differentiation and the differential time $\Delta t = 1.0 \ \mathrm{s}$. This model ship is equipped with three GNSS receivers to ensure redundancy. However, after confirming the positioning accuracy with the method described above, we used the GNSS receivers measurements only at the front of the model ship.

 The heading angle $\psi$ was calculated from the two GNSS receivers' relative positions at the front and rear of the model ship. This is because the drift of the FOG was non-negligible; approximately $5^{\circ}$ per 10 minutes during the measurement. The angular velocity $r$ is measured by the FOG and filtered by a low-pass filter with a cutoff frequency of 0.2 Hz.
 
 On the wind measurement, environment parameter  $\boldsymbol{\omega}$, which is the wind disturbance on this study, is obviously function of 3 dimensional space and time: $\boldsymbol{\omega}(x_{0},~y_{0},~z_{0},~t)$. However, measurement of high resolution space distribution at the experimental pond facility with was not practical from cost perspective. Hence, $\boldsymbol{\omega}$ was modeled as the function of time:
 $\boldsymbol{\omega}(t) = (U_{\mathrm{T}}(t), \ \gamma_{\mathrm{T}}(t))^{\mathrm{T}}$, and derived from the apparent wind velocity measured by two anemometers on the ship as follows. Measured apparent wind on each anemometers $(U_{A,k}(t) \ \gamma_{A,k}(t))$ were once converted to local true wind $\boldsymbol{\omega}_{k}(t) = (U_{\mathrm{T},k}(t), \ \gamma_{\mathrm{T},k}(t))^{\mathrm{T}}$. Then, the environment parameter  $\boldsymbol{\omega}(t)$ was derived by the average of local true wind to reduce the dependency to the space distribution:
 \begin{equation}
     \boldsymbol{\omega}(t) = \frac{1}{N}\sum_{k}\boldsymbol{\omega_{k}}(t) \enspace,
 \end{equation}
 where $N$ is the number of anemometers.

\subsection{Training and test data}
This section describes the detail of data sets used in this study as training data and test data. Training data is the input for the optimization process, and test data is another set of data to test the generalization performance of the optimal mathematical model. \Cref{tab:inputdata} shows the data set used in this study. In the table, the annotation R, T, Z, B-S, and B-P means the type of maneuver: R for random; T for turning; Z for zigzag; B-S and B-P for berthing to the starboard side and port side. Three sets of training data were used, which have different combinations of maneuver; Train-R, Train-TR, and Train-TZR, which contains random maneuver, random and turning, random, turning, and zigzag maneuver, respectively. These three data sets have an approximately equivalent duration of time. To set equivalent length, measured data were divided into several subsets. The annotation of the subsets "Train-R1" means the first subset of random maneuver.
\begin{table}[tb]
    \centering
    \caption{List of training and test data. Percentage inside () means the fraction of duration of each subset.}
    \begin{tabular}{llc}
    \hline  Data set & Subsets & Amount of data \\
    \hline Train-R & Train-R1 $(75\%)$, Train-R2 $(25\%)$ & $2695\ (\mathrm{s})$ \\
    Train-TR & Train-R1 $(78\%)$,                   & $2580\ (\mathrm{s})$ \\
             & Train-T1 $(10\%)$, Train-T2 $(11\%)$ & \\
    Train-TZR & Train-R1$ (76\%)$, & $2660\ (\mathrm{s})$ \\
              & Train-Z $(14\%)$, Train-T1 $(10\%)$ &  \\
    Test      &  Test-R $(44\%)$,  & $1624\ (\mathrm{s})$ \\
               & Test-Z $(9\%)$, & \\
               & Test-T $(32\%)$, & \\
               & Test-B-S $(8\%)$,& \\
               & Test-B-P $(7\%)$& \\
    \hline
    \end{tabular}
    \label{tab:inputdata}
\end{table}
\begin{table}[tb]
    \centering
    \caption{Control Input of Turning and Zigzag test data}
    \begin{tabular}{llc}
    \hline subsets Name & $\delta \ (\text{deg.})$ & $n_{\mathrm{p}} \ (\text{rps})$ \\
    \hline 
    Train-T1 & $-20$ &$10$ \\
    Train-T2 & $20$ &$10$ \\
    Train-Z & $15/15, \ 30/30$ & $10$ \\
    Test-T & $35$ & $8$ \\
    Test-Z & $20/20$ & $12$ \\
    \hline
    \end{tabular}
    \label{tab:tzb}
\end{table}

The random maneuver is a maneuver with random control inputs. The random maneuver aims to contain all possible values of control inputs $\boldsymbol{u}$ and states ${\boldsymbol{x}}$ to reduce necessary data for training and test. Additionally, by utilizing the random maneuver for training data, the obtained parameters will be more robust to a wide range of control input and state. Nonaka first introduced random input~\cite{Nonaka1972} on random rudder motion of free-running model test to estimate Abkowitz maneuvering model. The Pseudo random binary signal (PBRS)~\cite{Yoon2003} and multi-level pseudo-random signal (m-level PRS)~\cite{Wang2020b} are other kinds of random maneuvers for system identification that contain multiple: duration of certain rudder angles (PBRS); or amplitude of rudder angle (m-level PRS). We expand the idea of random input to both $\boldsymbol{u}$ and state ${\boldsymbol{x}}$ for efficient optimization of maneuvering model. Ideally, random control input must be truly random by predetermined inputs; however, the measurements in the pond have the risk of collision with the shore and grounding. Therefore, the control input was given by the shore-based operator's radio controller to make the distribution of the control inputs and state random as much as possible. The maximum and minimum of control inputs were $\delta \leq \pm 35^{\circ}$, $n_{\mathrm{p}}\leq \pm 10\ \textrm{rps}$. Cruising speed of 10 rps is equivalent to 7.7 knots at full scale when the ship reaches constant speed navigating straight forward at that $n_{\mathrm{p}}$ with scale.

In addition to the random maneuver, turning and zigzag maneuvers were used as training and test data. This is because: to add a portion of quasi-steady motion to the data set while the random maneuver is transient motion, the turning and zigzag maneuvers are very likely to be measured at sea trial. Hence those data are available for many ships. \Cref{tab:tzb} shows the control input of turning and zigzag maneuver. Those inputs were chosen not to overlap each other. This is because not use the same control input between training and test data. Note that the turning and zigzag maneuver data contains the course-keeping acceleration maneuver before the ship starts to turn or zigzag. The acceleration maneuver was included in the data set to train and test the ship's important feature, navigating straight forward under the wind disturbance. PD controller was used to maintaining heading during course-keeping maneuvers.

Since this research aims to establish an accurate maneuvering mathematical model applicable to berthing, we included berthing maneuver in the test data. The berthing maneuver was conducted in the center of the pond without the berth wall and controlled manually by the operator.  . The maximum and minimum of control inputs are $\delta \leq \pm 35^{\circ}$, $n_{\mathrm{p}}\leq \pm 20\ \textrm{rps}$, which is higher rps than random maneuvers. This is because the model ship \textit{Esso Osaka} could not control the berthing maneuver by the shore-based operator sufficiently.

\begin{figure}
    \centering
    \includegraphics[width=\linewidth]{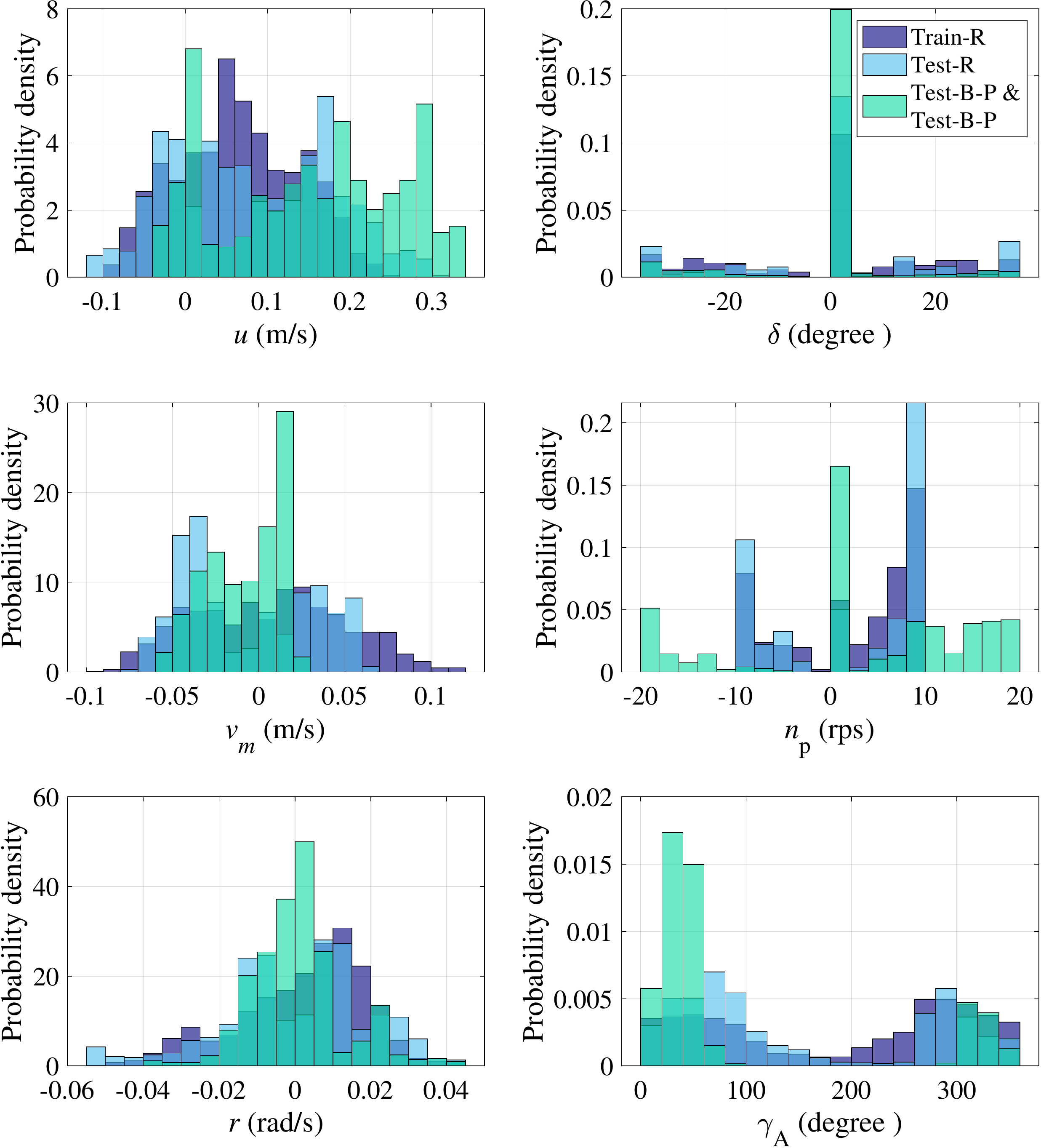}
    \caption{Histogram of state, control input and wind distribution of Train-R, Test-R, Test-B-S and Test-B-P data set.}
    \label{fig:PDF}
\end{figure}

The distribution of state and control input of random maneuver pretty much covers the berthing maneuver. \Cref{fig:PDF} shows the probability distribution function of Train-R, Test-R, and Test-B-S+Test-B-P data sets. Distribution of control inputs of random maneuver were biased around zero and the limit, $n_{\mathrm{p}} =0,\ \pm 10\ \mathrm{rps}$ and $\delta = 0, \ \pm 35^{^\circ}$, although the operator of model ship tried to make the control input as random as possible. Meanwhile, on the berthing maneuver data, larger revolutions numbers were used than the limit of $n_{\mathrm{p}}$ of random maneuver, and coasting $n_{\mathrm{p}} =0$ was more frequently used. On the distribution of $u$, $v_{m}$, and $r$, both random maneuver data set are generally well distributed among the range of slow-speed region, which is used on berthing maneuver data sets: Test-B-S and Test-B-P. The upper limit of $u$ was around $0.3<u<0.34 \mathrm{m/s}$, which is approximately 6 to 6.8 knot at full scale; a typical approaching speed of berthing. Apparent wind direction $\gamma_{\mathrm{A}}$ lucks the data on backward wind $90^{\circ}<\gamma_{\mathrm{A}}<270^{\circ}$. This was caused by the limitation of true wind direction $\gamma_{\mathsf{T}}$ due to the surrounding building of the pond. Additionally, Test-B-S and Test-B-P had more biased $\gamma_{\mathrm{A}}$ distribution because the initial heading of the berthing maneuver was fixed.

\section{Results}
In this section, to find the appropriate way to optimize the mathematical model, three objective functions $J_{1}, \ J_{2}, \ J_{3} $ defined in \cref{sec:J}, and three training data were compared. Once the best objective function and type of training data were selected, $\boldsymbol{x}(t)$ computed by the optimal mathematical model were compared with free-run model tests to evaluate the accuracy improvement by the proposed method. Additionally, simulations with the EFD model were compared to those with optimal parameters. 

\begin{figure}[tb]
    \centering
    \includegraphics[width=0.75\linewidth]{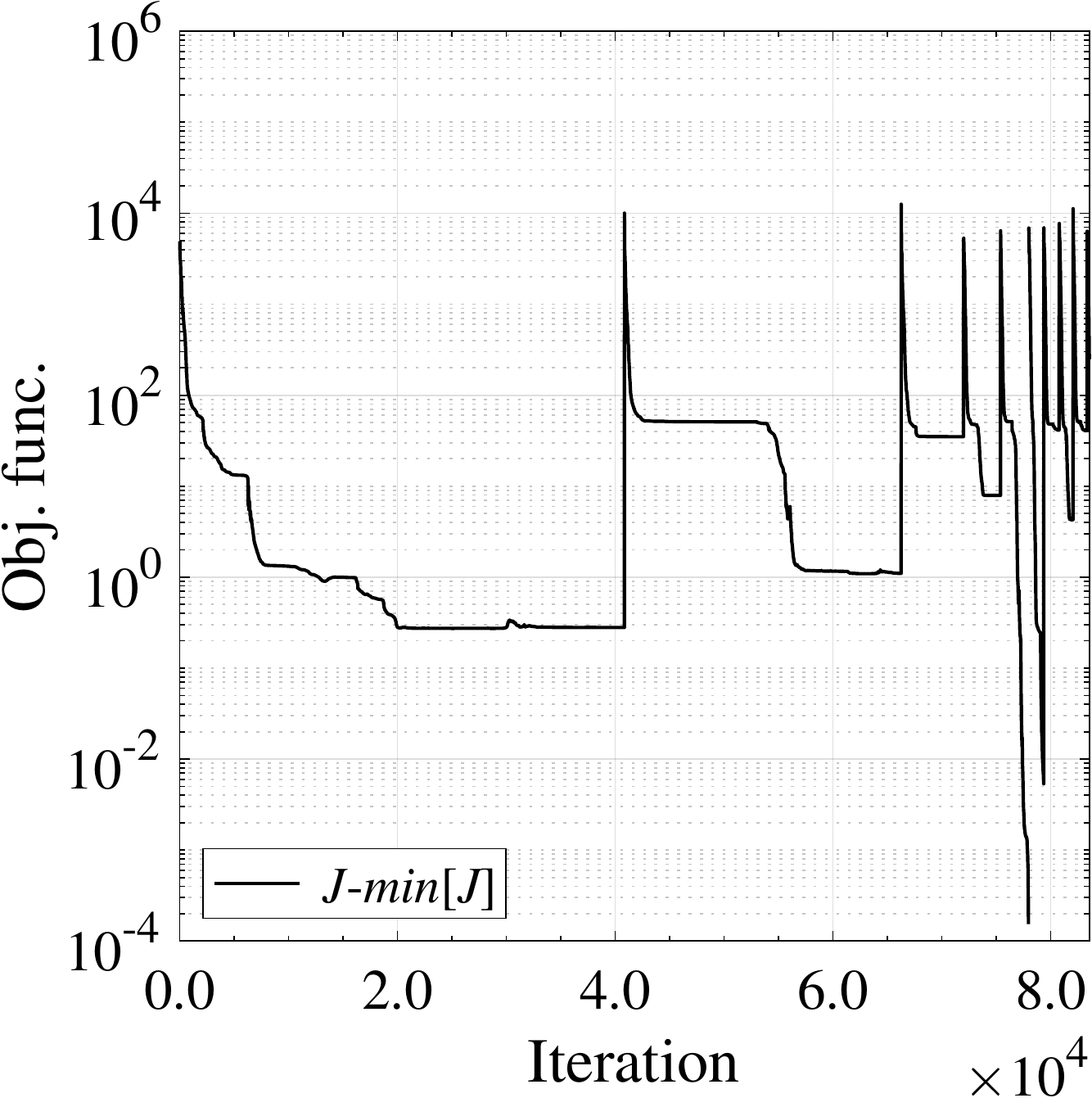}
    \caption{Optimization process by CMA-ES. The difference of objective function $J$ at each iteration and $\min(J)$.}
    \label{fig:conv}
\end{figure}

\paragraph{Convergence of Computation}
The iterative process in the optimization by CMA-ES is shown in \Cref{fig:conv}. The computation conditions of \Cref{fig:conv} were Train-R data set for input data and $J_{1}$ for objective function on the optimization process. The figure shows the difference of $J$ at each iteration and the minimum value of $J$ through the optimization process. On \Cref{fig:conv}, the iterative process shows impulse-like increases caused by the restart of CMA-ES. By using the restart strategy, CMA-ES lets the $J$ converge to several different local minima and choose the best solution from those. From \Cref{fig:conv}, we can see that the optimum solution $\boldsymbol{\theta}_{\text{opt}}$ is obtained at the 77970th iteration. In this study, the iteration in the optimization process continued until either it reached 100,000 iterations or a computation time of 5 days. Within the computation time of 5 days, the population size of CMA-ES reached to maximum population size in all cases.  All computation was conducted on the workstation equipped with Intel Xeon Gold 6248R for CPU.

\subsection{Comparison of Computation Results by Test Data Set}

This section shows the optimization by three training data sets and three objective functions to find the appropriate optimization method. Since the CMA-ES uses a stochastic approach, the result could depend on the random-seed. Hence three independent trials of optimization were conducted for each case. Note that with certain random-seed, the mean value of the population of CMA-ES does not converge to the domain of box constrain $\Theta_{j}$. Those results were removed from the random-seed trail if one of the obtained optimization target parameter $\theta_{j}$ is ten times larger than the boundary of $\Theta_{j}$. If the $\theta_{j}$ is smaller than the boundary of $\Theta_{j}$ times 10, obtained $\theta_{j}$ was used without correction to fit $\Theta_{j}$.
\Cref{tab:resultJ2_test} shows the comparison of $J_{2}$ on test data sets.

Objective functions $J_{1},~J_{2}$, and $J_{3}$ on test data set were used to evaluate the performance of optimization. $J_{2}$ served as the primary performance index because it contains both location and velocity component, those necessary to be estimated accurately on berthing. In addition to those physical meaning, $J_{2}$-Represents overall performance because $J_{2} = J_{1} + J_{3}$, as shown on \Cref{eq:z1,eq:z2,eq:z3}. \Cref{tab:resultJ1_test,tab:resultJ2_test,tab:resultJ3_test} shows the $J_{1},~J_{2}$, and $J_{3}$ on test data set.

From the sum of $J_{1},~J_{2}$, and $J_{3}$ on the overall test data shown on the  \Cref{tab:resultJ1_test,tab:resultJ2_test,tab:resultJ3_test},  we can analyze which combination of training data and the objective function is suitable in general for optimization. All optimal mathematical model gives lower value on $J_{1},~J_{2}$, and $J_{3}$ for overall test data than EFD model.
This means the present study's optimization method can generally improve estimation accuracy compared to the EFD model. Regarding the choice of the objective function and training data, the top three cases are $J_{2}$-R, $J_{1}$-R, $J_{1}$-TR for the evaluation by $J_{1}$; $J_{2}$-R, $J_{2}$-TR, $J_{1}$-R for evaluations by $J_{2}$ and $J_{3}$. $J_{2}$-R is the best condition on all three evaluation methods; however, the difference with second best, $J_{1}$-TR and $J_{1}$-R, were small compare to its standard deviation. Optimization using $J_{3}$ is worse than the other two, even on the evaluation by $J_{3}$ (\Cref{tab:resultJ3_test}), but no clear trend is shown between $J_{1}$ and $J_{2}$. 

Since the main objective of the present study is to develop a mathematical model that could accurately estimate the berthing maneuver, we find that this can be achieved by optimization using random ship maneuvers. Again from \Cref{tab:resultJ1_test,tab:resultJ2_test,tab:resultJ3_test}, we can see the performance of proposed method on berthing maneuver, by referring  the sum of $J_{1},~J_{2}$, or $J_{3}$ on Test-B-S and Test-B-P. Optimal models which used $J_{1}$ or $J_{2}$ on the optimization show better performance than the EFD model on all cases; however, optimization which used $J_{3}$ shows degraded performance. The top three cases are $J_{2}$-TR, $J_{2}$-R, $J_{1}$-R on all three evaluation methods. Same as on the overall test data, $J_{2}$-R and $J_{2}$-TR can estimate berthing maneuver but like overall test data, but those differences are small compare to its standard deviation.

Hence we can summarize the result on comparison to find the appropriate way of optimization: (i) optimization using $J_{2}$ with the data set of random maneuvering only (Train-R) and random maneuvering and turning (Train-TR) are the best choices for the estimation of berthing motion, random motion and overall test data; (ii) the difference between optimization with $J_{2}$-R and $J_{2}$-TR is small. Thus, we can not define whether Train-R or Train-TR is the best. This is because of the dominance of random maneuver on Train-TR data set, only 21\% of data is turning test (see \Cref{tab:inputdata}); (iii) optimization using $J_{3}$ has degraded performance compare to $J_{1}$ and $J_{2}$.  

\begin{table*}[tb]
    \centering
    \caption{Average of $J_{1}$ of three random-seed trails of each case on test data. The Values in \textbf{Bold} font are the best case for each test data; \underline{underlined} are the second-best; () are the standard deviation of random-seed trail. Case ``EFD'' means the simulation with EFD model. Other names of cases in the table represent the objective function and training data used in the optimization process; case $J_{1}$-R used the $J_{1}$ objective function and Train-R data set.}
    \label{tab:resultJ1_test}
    \begin{tabular}{c|rrrrrr|r}
    \hline Case & Test-B-S & Test-B-P &Test-B-S + Test-B-P& Test-R & Test-T & Test-Z & Total \\
\hline \hline  EFD & $251.9$ & $44.2$ & $296.1$ & $147.6$ & $843.8$ & $153.6$ & $1441.1$ \\
$J_{1}$-R & $144.4(24.9)$ & $\mathbf{55.2(1.6)}$ & $199.6(23.5)$ & $\mathbf{66.5(6.7)}$ & $\underline{167.4(35.1)}$ & $\underline{64.1(7.6)}$ & $\underline{497.7(69.4)}$ \\
$J_{1}$-TR & $126.7(6.2)$ & $78.1(1.3)$ & $204.8(7.5)$ & $88.2(16.4)$ & $\mathbf{156.6(12.7)}$ & $71.4(8.7)$ & $521.0(45.3)$ \\
$J_{1}$-TZR & $146.2(22.6)$ & $76.8(10.8)$ & $222.9(23.9)$ & $80.6(9.1)$ & $217.9(15.0)$ & $\mathbf{57.3(6.6)}$ & $578.7(15.0)$ \\
$J_{2}$-R & $\underline{108.5(9.1)}$ & $58.8(4.0)$ & $\underline{167.3(12.3)}$ & $\underline{74.3(6.7)}$ & $183.7(41.8)$ & $67.9(7.3)$ & $\mathbf{493.2(64.0)}$ \\
$J_{2}$-TR & $\mathbf{99.4(11.9)}$ & $64.6(3.5)$ & $\mathbf{164.0(8.8)}$ & $80.3(7.1)$ & $229.2(30.2)$ & $65.0(1.7)$ & $538.5(29.3)$ \\
$J_{1}$-TZR & $174.3(5.2)$ & $65.7(4.4)$ & $240.0(3.2)$ & $74.8(0.6)$ & $208.6(5.2)$ & $68.8(1.0)$ & $592.2(8.5)$ \\
$J_{3}$-R & $154.2(41.3)$ & $85.4(1.0)$ & $239.5(42.3)$ & $144.9(54.0)$ & $256.6(141.8)$ & $74.2(21.2)$ & $715.2(112.2)$ \\
$J_{3}$-TR & $281.0(24.6)$ & $\underline{57.7(10.3)}$ & $338.7(24.9)$ & $220.6(58.4)$ & $586.5(205.6)$ & $69.9(5.3)$ & $1215.7(245.6)$ \\
$J_{3}$-TZR & $270.2(42.5)$ & $312.5(206.2)$ & $582.7(204.8)$ & $236.1(127.5)$ & $233.3(34.6)$ & $92.1(20.8)$ & $1144.3(376.5)$ \\
\hline
\end{tabular}
\end{table*}
\begin{table*}[tb]
    \caption{Average of $J_{2}$ of three random-seed trails of each case on test data.  Notations of this table are the same as \Cref{tab:resultJ1_test}.
    }
    \centering
    \begin{tabular}{c|rrrrrr|r}
    \hline Case & Test-B-S & Test-B-P &Test-B-S + Test-B-P& Test-R & Test-T & Test-Z & Total \\
    \hline \hline  EFD & $737.2$ & $453.1$ & $1190.3$ & $478.8$ & $1259.5$ & $709.3$ & $3637.9$ \\
    $J_{1}$-R & $485.0(70.6)$ & $\mathbf{183.3(13.8)}$ & $668.4 (59.4)$ & $\underline{159.5(27.3)}$ & $\underline{253.0(84.1)}$ & $597.5($ 7.7 $)$ & $1678.3(166.9)$ \\
    $J_{1}$-TR & $431.9(22.1)$ & $387.4(40.6)$ & $819.4(18.9)$ & $206.9(12.1)$ & $\mathbf{207.4(1.0)}$ & $608.7(82.5)$ & $1842.3(76.8)$ \\
    $J_{1}$-TZR & $428.4(112.0)$ & $331.1(82.1)$ & $795.5(64.0)$& $176.9(22.2)$ & $339.0(51.5)$ & $583.7(33.0)$ & $1859.0(114.4)$ \\
    $J_{2}$-R & $\underline{359.9(68.3)}$ & $\underline{224.9 (35.6)}$ & $\underline{584.8 (84.7)}$ & $\mathbf{154.6(12.4)}$ & $259.8(84.8)$ & $605.5(16.0)$ & $\mathbf{1604.7(191.8)}$ \\
    $J_{2}$-TR & $\mathbf{310.0(37.9)}$ & $258.7(50.0)$ & $\mathbf{568.8 (12.7)}$ & $178.3(17.6)$ & $308.5(23.7)$ & $604.2(4.5)$ & $\underline{1659.7(36.4)}$ \\
    $J_{2}$-TZR & $523.8(13.5)$ & $266.8(72.9)$ & $790.0 (61.4)$ &$206.9(3.1)$ & $274.0(9.8)$ & $\underline{555.7(3.3)}$ & $1827.2(70.2)$ \\
    $J_{3}$-R & $415.5(182.4)$ & $392.2(43.8)$ & $807.7(209.2)$& $300.9(44.8)$ & $335.3(129.6)$ & $561.3(61.6)$ & $2005.2(141.0)$ \\
    $J_{3}$-TR & $647.7(68.2)$ & $332.6(119.5)$ & $980.3 (98.4)$ & $432.8(102.7)$ & $1049.1(562.2)$ & $\mathbf{500.3(101.2)}$ & $2962.5(457.5)$ \\
    $J_{3}$-TZR & $677.9(123.1)$ & $791.4(319.8)$ & $1469.3(329.4)$ & $534.0(290.1)$ & $301.0(53.5)$ & $650.6(76.5)$ & $2955.0(733.8)$ \\
    \hline
    \end{tabular}
    \label{tab:resultJ2_test}
\end{table*}
\begin{table*}[tb]
    \centering
    \caption{Average of $J_{3}$ of three random-seed trails of each case on test data. Notations of this table are the same as \Cref{tab:resultJ1_test}.}
    \label{tab:resultJ3_test}
    \begin{tabular}{c|rrrrrr|r}
    \hline Case & Test-B-S & Test-B-P &Test-B-S + Test-B-P& Test-R & Test-T & Test-Z & Total \\
\hline \hline  EFD & $485.3$ & $408.9$ & $894.2$ & $331.2$ & $415.7$ & $555.7$ & $2196.9$ \\
$J_{1}$-R & $340.6(45.6)$ & $\mathbf{128.2(13.0)}$ & $468.8(36.5)$ & $\underline{92.9(21.7)}$ & $85.5(49.2)$ & $533.4(11.5)$ & $1180.6(97.6)$ \\
$J_{1}$-TR& $305.3(15.9)$ & $309.3(41.9)$ & $614.6(26.3)$ & $118.7(4.4)$ & $\mathbf{50.8(11.8)}$ & $537.3(73.7)$ & $1321.4(31.5)$ \\
$J_{1}$-TZR & $282.2(93.4)$ & $254.3(72.0)$ & $536.5(40.9)$ & $96.3(17.6)$ & $121.1(66.5)$ & $526.4(36.8)$ & $1280.3(111.2)$ \\
$J_{2}$-R& $\underline{251.4(59.4)}$ & $\underline{166.1(31.8)}$ & $\underline{417.5(72.4)}$ & $\mathbf{80.3(8.2)}$ & $76.1(43.8)$ & $537.6(8.9)$ & $\mathbf{1111.5(131.5)}$ \\
$J_{2}$-TR& $210.6(28.4)$ & $194.1(46.6)$ & $\mathbf{404.7(18.6)}$ & $98.0(11.0)$ & $79.3(6.5)$ & $539.2(4.6)$ & $\underline{1121.2(22.6)}$ \\
$J_{2}$-TZR & $349.5(8.4)$ & $201.1(68.4)$ & $550.6(60.6)$ & $132.1(2.7)$ & $\underline{65.4(5.8)}$ & $\underline{487.0(2.8)}$ & $1235.0(66.3)$ \\
$J_{3}$-R & $261.3(149.7)$ & $306.8(44.2)$ & $568.1(181.4)$ & $156.1(10.8)$ & $78.8(15.7)$ & $487.1(45.8)$ & $1290.0(235.4)$ \\
$J_{3}$-TR & $366.6(46.1)$ & $274.9(124.1)$ & $641.6(117.3)$ & $212.2(45.4)$ & $462.6(359.2)$ & $\mathbf{430.4(99.2)}$ & $1746.7(227.9)$ \\
$J_{3}$-TZR & $407.7(81.5)$ & $479.0(118.9)$ & $886.7(125.0)$ & $297.8(162.8)$ & $67.7(24.6)$ & $558.5(56.1)$ & $1810.7(359.4)$ \\
\hline
\end{tabular}
\end{table*}

\begin{figure*}[tb]
    \centering
    \includegraphics[width=\linewidth]{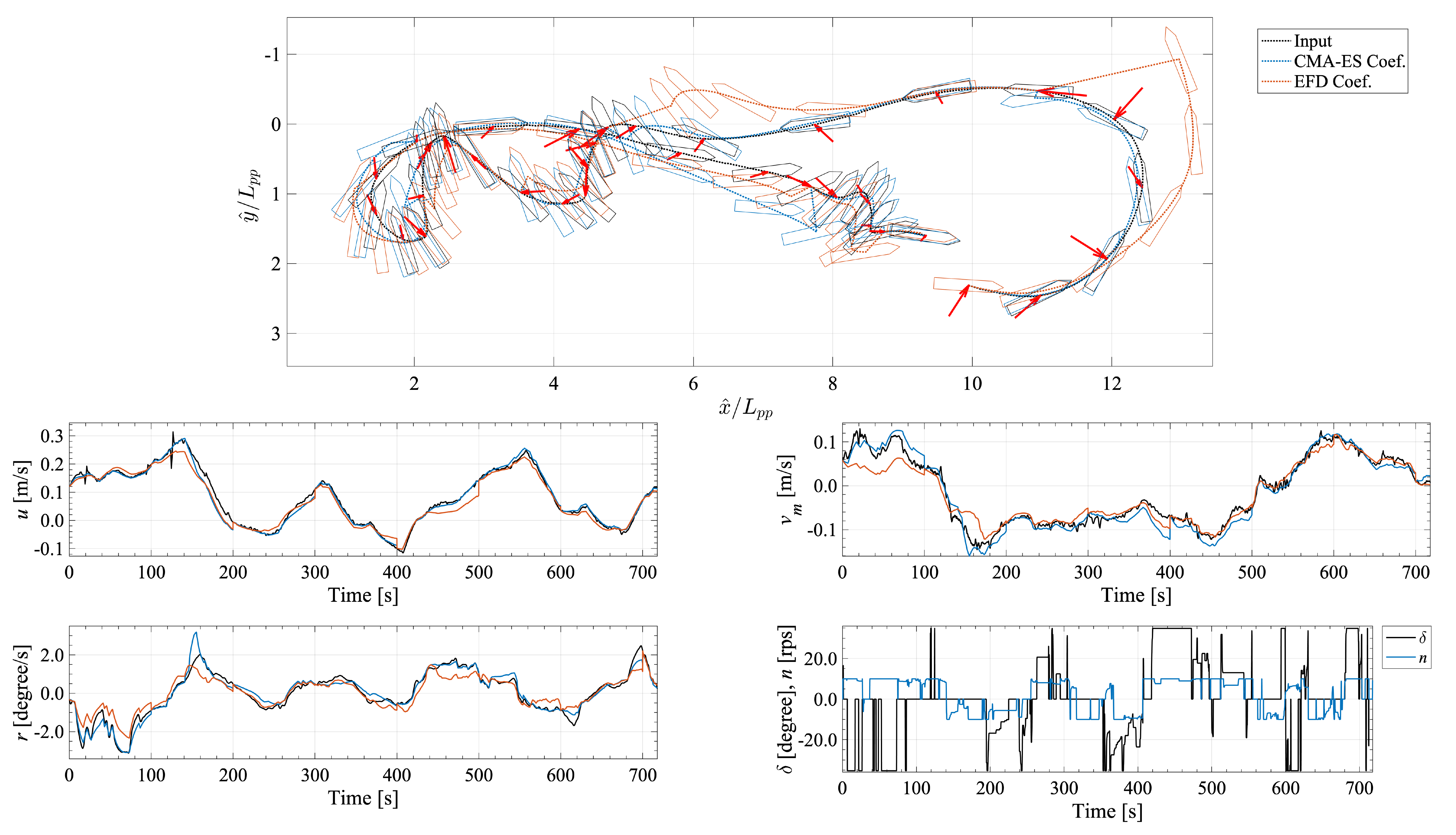}
    \caption{Estimated state of random maneuver test data using coefficients obtained by EFD and CMA-ES. Comparison with Input data: measured by free-run model test. The ship locations on upper figure only shown on $t(0), \ t_{f}^{i}$, and every 20 seconds.}
    \label{fig:test_random}
\end{figure*}

Finally, we show the comparison of $\boldsymbol{x}_{\text{sim}}(t)$ between the EFD model and the best case of the optimal mathematical model; one of the results of the random-seed trial of $J_{2}$-R. \Cref{fig:test_random} shows the $\boldsymbol{x}_{\text{sim}}(t)$ and $\boldsymbol{x}_{\text{input}}(t)$ on Test-R data set. Because the measured trajectories were divided into several contiguous subsequences, the path of ship ``jumps'' and time series of $u,\ v_{m}, \ r$ show the discontinuity at the end of contiguous subsequence. The red arrow in \Cref{fig:test_random} shows the apparent wind velocity at midship of the free-run model test result. From the figure, we can see the path and state computed by the optimal model agree well with the free-run model test result than the EFD model, even on the complex random maneuver at the outdoor environment with wind disturbance. Moreover, the Test-R data set contains well-distributed state and control input. This means the optimal model is numerically stable to the practical range of input state and control input of berthing maneuver.

\begin{figure*}[tb]
    \centering
    \includegraphics[width=\linewidth]{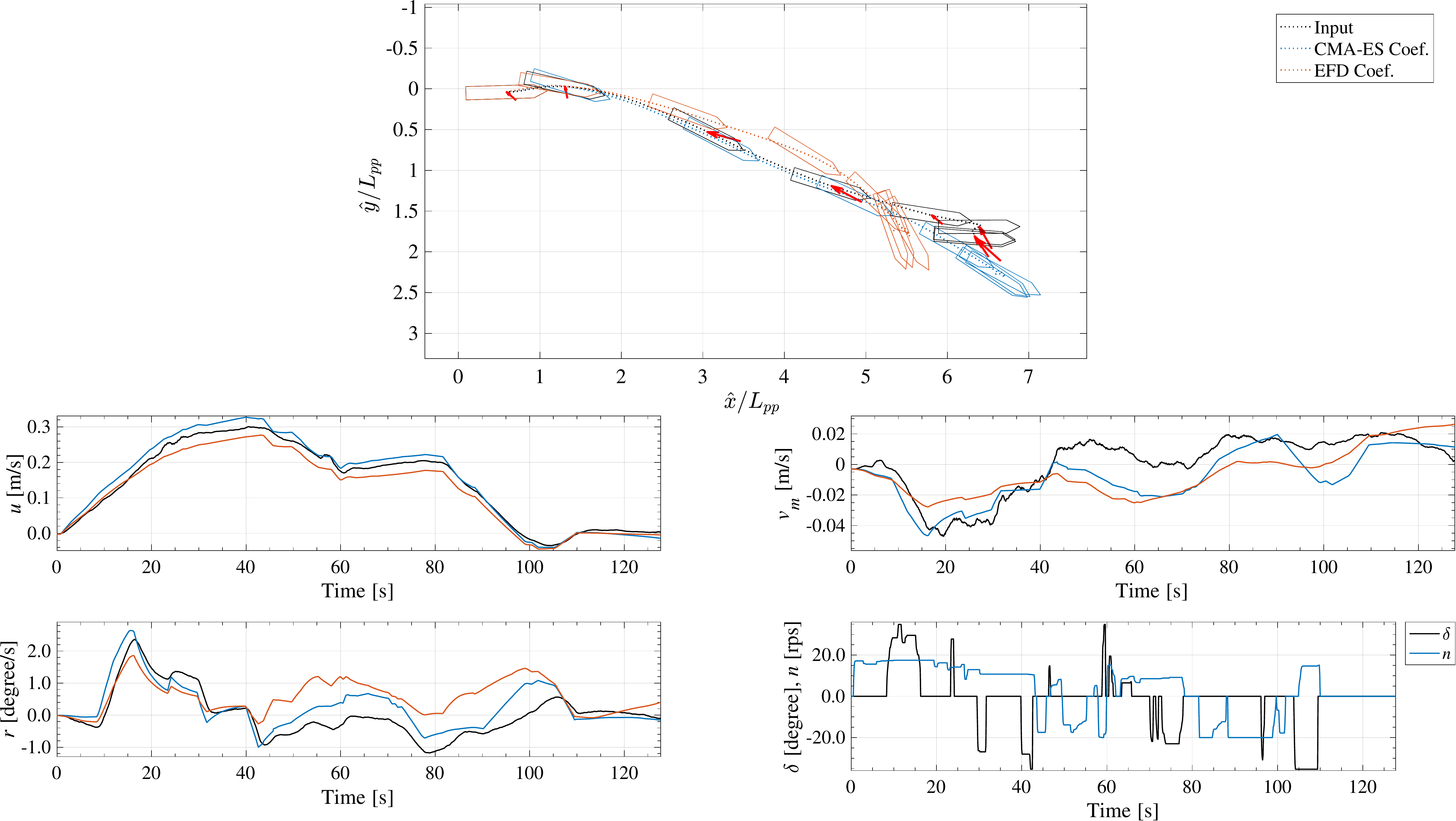}
    \caption{Estimated state of Starboard side berthing maneuver test data using coefficients obtained by EFD and CMA-ES. Comparison with Input data: measured by free-run model test.}
    \label{fig:test_berthS}
\end{figure*}

\begin{figure*}[tb]
    \centering
    \includegraphics[width=\linewidth]{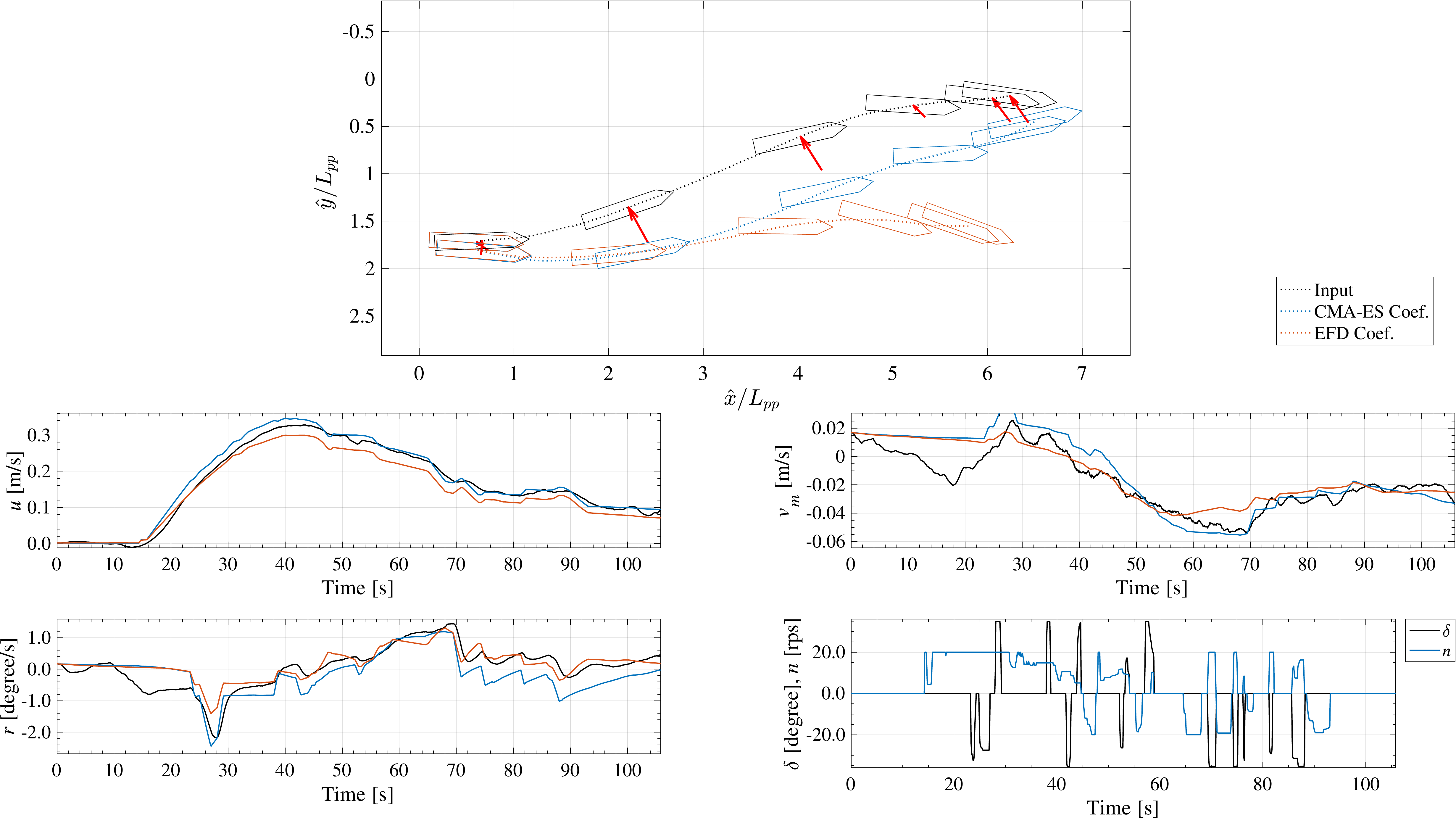}
    \caption{Estimated state of Port side berthing maneuver test data using coefficients obtained by EFD and CMA-ES. Comparison with Input data: measured by free-run model test.}
    \label{fig:test_berthP}
\end{figure*}

Comparison on berthing maneuver test data, Test-B-S and Test-B-P are shown on \Cref{fig:test_berthS,fig:test_berthP}. Note that input data was not divided into the contiguous subsequence on those comparisons because the duration of the free-run model test was nearly equal to the length of the contiguous subsequence. Although the $\boldsymbol{x}_{\text{sim}}(t)$ obtained by the optimal model not shows significant agreement with free-run model test result as on Test-R, however, the optimal model shows better $J_{2}$ on both berthing data set to compare to EFD model as shown on \Cref{tab:resultJ2_test}, and it should be emphasized that even berthing maneuver was not included to training data. Additionally, as shown on \Cref{fig:PDF}, even the range of propeller rev. of test data $n_{p} \leq \pm 20$ was exceeded the limit of training data $n_{p} \leq \pm 10$, 
the optimal model proposed in this study did not be numerically unsteady.

\section{Discussion}
The results shown in the previous section shows that the proposed method is feasible and practical to construct the berthing maneuver capable mathematical model by system parameter exploration from physically (not numerically) obtained, noisy trajectories. The features of the proposed method are: use model-rich MMG model to express the complex characteristics of berthing maneuver; use random maneuvers as training and test data. Optimal parameters estimated the ship's maneuver better than the CMT-based scheme, the berthing maneuver, the complex random maneuvers, and typical maneuvers: turning and zig-zag. Hence, the mathematical model with optimal parameters is applicable for both maneuverings inside and outside the harbor.

The advantages of the proposed method are as follows: (1) It requires a smaller amount of data to construct a mathematical model compared to the CMT-based scheme. The CMT requires a great number of model tests to obtain the whole list of system parameters, while the proposed method only needs approximately one hour of the trajectory of the free-run model test for training and test data. This point becomes obvious for berthing maneuver application because to achieve sufficient accuracy of the MMG model for berthing maneuver, the number of system parameters increases rapidly, which requires more test conditions for the CMT. (2) Able to overcome the scale effect by using the trajectory of the full-scale ship. Because the CMT-based scheme has no established way to correct the scale effect,  the mathematical model needs to be tuned manually when it is applied to the full-scale ship, as stated in \Cref{sec:intro}. 
Although the free-run model test trajectories were used, the proposed method provides flexibility to obtain the optimal system parameter from the trajectories measured on a full-scale ship. This implies that the scale effect is irrelevant when constructing the mathematical model.
(3) The domain of the exploration is wide enough to be independent of the EFD result, which defines the domain boundary. Because the typical measurement error and variation of the parameter between different ships are much smaller than the exploration range (i.e., x10 of reference EFD obtained parameter), the proposed method can obtain an adequate mathematical model even if the EFD result was obtained for another ship (e.g., $k_{xOR}, \ C_{PR}$). This is preferable when the EFD result is not available for the newly designed ship.

The major drawbacks of the proposed method are as follows. 
First, the modular structure, which is the advantage of the MMG model, is compromised. Once the ship's design is modified, the CMT-based scheme only needs to reacquire the parameter related to the modification, while the proposed method needs to reacquire the whole data set. Additionally, because the parameter will drift when several parameters are estimated simultaneously (this effect is known as the ``cancellation effect''~\cite{Hwang1982}), the obtained parameter by the proposed method can not be used independently of other parameters.
Second, random-seed trials are necessary because the variation of the stochastic search was relatively large.

The remaining issues of the proposed method are consideration on: limitation of mathematical model's degree of freedom; and dependency on the amount of training data. First, although the mathematical model used in this study includes the most sophisticated and complicated MMG sub-models, still unable to capture the complex hydrodynamic phenomena in a berthing maneuver completely. The MMG model of this study assumes several parameters as constant even if they vary on a large drift angle; for instance, rudder force increase factor $a_{H}$ varies and changes its sign when drift angle  $|\beta|\geq 45^{\circ}$~\cite{Yasukawa2021}. The performance of SI relies on the mathematical model's degree of freedom. Hence, the mathematical model needs to be improved to enhance the capability of the present method. Second, the dependency on the amount of training must be investigated. In this study, we obtained the mathematical model with satisfactory accuracy, though the amount of training data was approximately constant. The amount of data was limited to approximately seven hours on the full scale for practical use. However, for future work, the possibility of accuracy improvement by adding the data must be investigated. 

Although the remaining issues are stated above, the proposed method will be one of the practical schemes to obtain an accurate mathematical model to estimate berthing maneuver with relatively low computational cost. Additionally, as stated in~\cite{Araki2012a}, combining SI with the direct CFD estimation could construct a reliable mathematical model without any model test or sea trial. This will make it easier to adjust the berthing control algorithm during the ship design phase.

\section{conclusion}
Accurate maneuvering estimation is essential to establish autonomous berthing control, critical technologies for autonomous shipping. The system-based mathematical model is widely used to estimate the maneuver. Commonly, the system parameters of the model are obtained by the CMT, which is time-consuming to construct an accurate model suitable for complex berthing maneuvers.
System identification (SI) is one option to construct the maneuvering model, which requires only a few trajectories for training data; however, SI on a mathematical model of ship's maneuver was only conducted on much simpler maneuver, turning and zig-zag, in the past. 

This study investigated the feasibility of SI of a system-based method to establish an accurate estimation method of berthing maneuver for autonomous berthing control.
The SI of the MMG model with a global optimization scheme can obtain a reasonable mathematical model to represent the dynamics of complex berthing maneuvers with relatively few number of trajectories data for training. The simulation using optimal parameters showed better agreement with the free-running model test than the CMT-based scheme. We found that using both position and velocity components on the objective function with the data set of random maneuvering only (Train-R) or random maneuvering and turning (Train-TR) are the best choices for optimization. The proposed method can obtain the mathematical model for berthing maneuver and can be one of the promising alternatives to the CMT-based scheme due to its reduced number of required model tests.

\begin{acknowledgements}
This study was supported by a Grant-in-Aid for Scientific Research from the Japan Society for Promotion of Science (JSPS KAKENHI Grant \#19K04858). The study also received assistance from the JFY2018 Fundamental Research Developing Association for Shipbuilding and Offshore (REDAS) in Japan. The authors also would like to express gratitude to Mr. Satoru Konishi, Magellan Systems Japan Inc., for the technical support on GNSS measurement during the free run model test. Finally, the authors would like to thank Nozomi Amano and Yuta Fueki, Osaka University, for supporting the free-run model test, and Koki Wakita, Osaka University, for technical discussion.
\end{acknowledgements}

%
\section*{Conflict of interest}

The authors declare that they have no conflict of interest.

\bibliographystyle{spphys}       
\bibliography{main.bib}   

\end{document}